\documentclass[useAMS,usenatbib]{mn2e}
\usepackage{pdfpages}
\usepackage{wasysym}
\usepackage{graphics}
\usepackage{graphicx}
\usepackage{color}
\usepackage{natbib}
\usepackage{aastex_hack}
\usepackage{times}
\usepackage{float}
\usepackage{color}

\usepackage{subfigure}

\title[Lambda Boo star debris disks]{IR-excesses around nearby Lambda Boo stars are caused by debris disks rather than ISM bow waves}
\author[Z. H. Draper et al.]{Z. H. Draper$^{1,2}$\thanks{E-mail: zhd@uvic.ca}, B. C. Matthews$^{2,1}$, G. M. Kennedy$^{3}$, M. C. Wyatt$^{3}$, K. A. Venn$^{1}$, B. Sibthorpe$^{4}$ \\
$^{1}$Department of Physics and Astronomy, University of Victoria, 3800 Finnerty Rd, Victoria, BC V8P 5C2, Canada\\
$^{2}$Herzberg Institute of Astrophysics, National Research Council of Canada, 5071 West Saanich Road., Victoria, BC V9E 2E7, Canada\\
$^{3}$Institute of Astronomy, University of Cambridge, Madingley Road, Cambridge CB3 0HA, UK\\
$^{4}$SRON Netherlands Institute for Space Research, Landleven 12, NL-9747 AD Groningen, the Netherlands}
\begin{document}

\date{Accepted 2015 November 16  Submitted 2015 June 29}

\pagerange{\pageref{firstpage}--\pageref{lastpage}} \pubyear{2002}

\maketitle

\label{firstpage}

\begin{abstract}
{Lambda Boo stars are predominately A-type stars with solar abundant C, N, O, and S, but up to 2 dex underabundances of refractory elements.  The stars' unusual surface abundances could be due to a selective accretion of volatile gas over dust.  It has been proposed that there is a correlation between the Lambda Boo phenomenon and IR-excesses which are the result of a debris disk or interstellar medium (ISM) interaction providing the accreting material. We observe 70 or 100 and 160 $\mu$m excess emission around 9 confirmed Lambda Boo stars with the \textit{Herschel Space Observatory}, to differentiate whether the dust emission is from a debris disk or an ISM bow wave. We find that 3/9 stars observed host well resolved debris disks. While the remaining 6/9 are not resolved, they are inconsistent with an ISM bow wave based on the dust emission being more compact for its temperature and predicted bow wave models produce hotter emission than what is observed.  We find the incidence of bright IR-excesses around Lambda Boo stars is higher than normal A-stars.  To explain this given our observations, we explore Poynting-Robertson (PR) drag as a mechanism of accretion from a debris disk but find it insufficient.  As an alternative, we propose the correlation is due to higher dynamical activity in the disks currently underway.  Large impacts of planetesimals or a higher influx of comets could provide enough volatile gas for accretion. Further study on the transport of circumstellar material in relation to the abundance anomalies are required to explain the phenomenon through external accretion.}
\end{abstract}


\section{Introduction}

The Lambda Boo stars are a class of Population I, B9-F3 type stars ($\sim$1.5-2.5 solar masses) of various ages with strongly depleted $\alpha$ and Fe-peak element abundances, but relatively normal solar abundances of C, N, O, and S \citep{EP04}.  The distinction between these species is that C, N, O, and S have a lower sublimation temperature on dust grains \citep{KL03}.  An abundance anomaly could be formed where volatile elements are accreted onto the star in the gas state while the refractory elements are locked away in dust grains which are blown away from the star due to radiation pressure \citep{VL90,LW92}.  The accretion would need to be relatively recent given the fact that meridional circulation would mix the surface in 1-2 Myr timescales \citep{ST02}.  While Lambda Boo stars are particularly metal deficient, A stars in the solar neighbourhood are typically found to be metal rich in Fe-peak elements, likely due to their post-solar formation age or correlation with the AmFm phenomenon \citep{GH95,SM12}. The abundance pattern based on sublimation temperature or excitation potential allows these stars to be distinguished from intrinsically metal weak stars such as Pop II or F-weak stars to constitute a class of their own \citep{EP14}. 

Multiple theories have been proposed to explain the Lambda Boo phenomenon, but none have been proven to be the direct cause of the surface abundance pattern.  There are 34 confirmed Lambda Boo stars \citep{GC93}, or approximately 50 including candidates, which suggests they are less than 2\% of all stars within their spectral range \citep{GC02}. This rarity requires the mechanism to be too weak to be widely observed or occur infrequently, but also explain their many unique properties. Theories for what causes the Lambda Boo phenomenon fall into two categories: those internal to the photosphere and those external to the photosphere.  

One of the internal mechanisms which has been proposed is a modification of the mass-loss theory in AmFm stars \citep{GM83}.  AmFm stars have a chemical peculiarity on their surface which has been observed and modelled by a selective diffusion of heavy metals towards the surface \citep{RJ00}.  If the mass-loss due to a radiatively driven stellar wind on an AmFm star was on the order of $10^{-13} M_{\astrosun}/\rm{yr}$, then it could produce a Lambda Boo-like signature on the surface because the radiation pressure would be more efficient for heavy metals \citep{GM83}. However, the mass-loss rates have not been found to be significant around AmFm stars \citep{RJ00}.  Furthermore, the AmFm phenomenon is not observed in stars with equatorial rotation speeds above $\sim$90 km/s due to meridional circulation mixing the abundances with the lower layers of the star \citep{PC93}.  Since Lambda Boo stars on average rotate with a $v \sin(i)$ of 120 km/s, this mechanism alone is not plausible to explain Lambda Boo-like abundances.

There are other observed effects that may point to an unknown internal mechanism. For instance, Lambda Boo stars are more likely to pulsate in the instability strip \citep{DB99,EP04}.  This is often called the $\delta$ Scuti phenomenon and is due to the increasing/decreasing of the opacity of the ionized helium boundary layer. This oscillation in ionizing helium occurs because the ionization temperature is within the internal temperature and pressure range of a main-sequence A star \citep{BS08}.  Lambda Boo stars are characterized by higher overtone modes rather than lower mode oscillations typical of $\delta$ Scuti stars \citep{EP04}.  However, there is no observed correlation with the Period-Luminosity-Colour relation of $\delta$ Scuti stars and metallicity, which would distinguish Lambda Boo  stars from ``normal'' $\delta$ Scuti stars \citep{EP02}. It may be that the pulsations are linked to a yet unknown diffusion or mixing process in main sequence A stars which causes the abundance pattern on the surface \citep{EP04}.  This has yet to be a well developed theory, but it is a unique characteristic of these stars which is rooted in observations and therefore should not be ignored \cite[e.g. see][]{MO10}

As for external mechanisms, spectroscopic binaries, debris disks, and ISM interactions have been proposed to be the cause.  In the case of close binaries, the Lambda Boo phenomenon may not be a real phenomenon, but is rather an artifact from not resolving the stars \citep{RF04}.  In this sense, the convolution of two stellar spectra make the A-star spectra seem metal poor when it actually isn't.  For example, one of the stars in our sample, HD 11413, was found to be a composite spectra binary via cross-correlation with a synthetic spectrum.  This method is prone to systematic error and was not a definitive radial velocity (RV) detection of a binary.  \cite{EG12} did a multi-year spectroscopic survey of Lambda Boo stars to detect RV shifts and found none to be composite spectra binaries.  Some of those stars are considered here, including HD 125162, HD 183324, and HD 221756.  This in general contradicts the claim of a composite spectra for HD 11413 causing its Lambda Boo-like properties.

The other external mechanisms considered are debris disks and ISM interactions. Both of these mechanisms superficially pollute the surface with gas but push out metal-rich grains via radiation pressure \citep{LW92}.  These two external mechanisms cannot be distinguished with spectral energy distribution (SED) characterization alone since they both result in thermal emission from dust, which manifests as excess flux density above the expected stellar photosphere.  Debris disks have been detected around $\sim$24\% of A stars \citep{NT14} with \textit{Herschel} and result from the collisional cascade of comets and asteroids which generate dust \citep{AC97}.  The ISM interaction will create a bow wave of locally heated dust in the direction of motion as the star passes through the cloud.  Typically, bow waves have been referred to as ``bow shocks" in the literature, but this implies that the gas is being shocked, which is not what is being modelled, so we will not refer to it as such.  Since the gas density can be low and dust does not shock like a fluid, the dust is rather pushed around a cone of avoidance from radiation pressure.  The significant fraction of the emission will result from the over density of warm dust in the bow front, but emission will also arise from the surrounding cloud especially when observed at longer wavelengths.

There are however some problems with the ISM accretion theory.  First, $\delta$ Velorum is a star which is well known to be interacting with the ISM, but is not a Lambda Boo star \citep{GA08}.  It has a well resolved, asymmetric bow structure seen with \textit{Spitzer} and was modelled with dust around 0.1 $\mu$m in size and astrosilicate composition, typical for the ISM.  Furthermore, most stars in the local solar neighbourhood, where some Lambda Boo stars actually reside (including all of the stars presented in this paper), have a low probability of interacting with the ISM. Although, other instances of ISM accretion can be found in more dense ISM regions exterior to the local bubble or in the galactic centre \citep{RB13}.  In those cases, polarimetry has been shown to aid in the identification of bow waves through the determination of the polarization angle relative to the emission, which helps differentiate the origin of the IR excess emission as bow waves.  However, AO assisted spectroscopy from large ground-based telescopes would be required to identify a Lambda Boo-like signature in these stars, which has not been done.  This could cement the relationship between ISM interactions and surface abundance anomalies. Other metal poor stars have also been searched for nearby excess emission to explain the stars' abundance anomalies. \cite{KV14} found no excess emission, but the observations were not sensitive enough to rule out a typical debris disk.  

Both of the accretion scenarios are supported by observations of gas toward several Lambda Boo stars \citep{HS93}. If gas is detected, then it may indicate a reservoir of volatiles is available to accrete on to the star. For example, spectroscopic shell line detections of Ca II or Na I are indicative of gas near the star \citep{DB94}.  However, the true location of the gas is unknown as the spectra are the culmination of material along the line of sight and therefore may not be associated with the star itself.  UV observations with higher energy transitions at RV shifts consistent with Keplerian motion provide confirmation that the gas is indeed circumstellar, such as around the Lambda Boo star 131 Tau \citep{CG96}.  However, this star was observed with \textit{Spitzer} and no excess dust emission was detected out to 70$\mu$m \citep{KS06}, so the source of this accreting gas is still unknown, but adds credence to external accretion mechanisms.  Time variable spectroscopic absorption, often interpreted as falling evaporating bodies (or FEB), have also been observed around Lambda Boo stars \citep{MW12}.  

It is also important to note that any gas accretion will be counter balanced by meridional circulation. This could negate a polluted abundance pattern in $\sim$10$^{6}$yrs and yet Lambda Boo stars are observed at various ages in their $\sim$Gyr main sequence lifetime \citep{ST02}.  This requires the Lambda Boo mechanism to operate at any time during the main sequence, which actually gives some additional support to the external mechanisms.  ISM interactions can occur randomly with age.  Debris disks form out of protostellar material and deplete with age, yet they have still been found around stars of $\sim$1 Gyr in age, subgiants, and even white dwarfs \citep{AB14,BM14}.  If the mechanism were internal, A stars would need to have a very specific criteria for initiating and ceasing a Lambda Boo-like phase, independent of age and stellar evolution. 

\begin{table}
\centering
\caption{List of Herschel observation ID numbers for target stars used in this paper. The RA and Dec listed are the observational centre pointings. \label{obsid}}
\begin{tabular}{cccc}
\hline
Star (HD) & RA & Dec & ObsID(s) \\
\hline
~~11413 & $\rm01^{h}49^{m}06^{s}\hspace{-1mm}.5$ & $\rm-50^{\circ}12\arcmin19\arcsec\hspace{-1mm}.8$ & 1342224383 / 84 \\
~~30422 & $\rm04^{h}46^{m}25^{s}\hspace{-1mm}.7$ & $\rm-28^{\circ}05\arcmin14\arcsec\hspace{-1mm}.6$ & 1342242078 / 79 \\
~~31295 & $\rm04^{h}45^{m}56^{s}\hspace{-1mm}.1$ & $\rm+10^{\circ}09\arcmin01\arcsec\hspace{-1mm}.4$ & 1342241872 / 73 \\
~~74873 & $\rm08^{h}46^{m}56^{s}\hspace{-1mm}.0$ & $\rm+12^{\circ}05\arcmin14\arcsec\hspace{-1mm}.6$ & 1342254577 / 78 \\
110411 & $\rm12^{h}41^{m}53^{s}\hspace{-1mm}.0$ & $\rm+10^{\circ}14\arcmin08\arcsec\hspace{-1mm}.2$ & 1342212660 / 61 \\
125162 & $\rm14^{h}16^{m}23^{s}\hspace{-1mm}.0$ & $\rm+46^{\circ}51\arcmin07\arcsec\hspace{-1mm}.9$ & 1342210928 / 29 \\
188324 & $\rm19^{h}29^{m}01^{s}\hspace{-1mm}.0$ & $\rm+01^{\circ}57\arcmin01\arcsec\hspace{-1mm}.8$ & 1342231678 / 79 \\
198160 & $\rm20^{h}51^{m}38^{s}\hspace{-1mm}.4$ & $\rm-62^{\circ}25\arcmin47\arcsec\hspace{-1mm}.0$ & 1342232490 / 91 \\
221756 & $\rm23^{h}34^{m}37^{s}\hspace{-1mm}.5$ & $\rm+40^{\circ}14\arcmin10\arcsec\hspace{-1mm}.3$ & 1342223973 / 74 \\
\hline
\end{tabular}
\end{table}

In order to place constraints on the two accretion mechanisms, we utilize \textit{Herschel} observations of known Lambda Boo stars with IR-excesses indicative of nearby dust to differentiate whether the IR-excess around nearby Lambda Boo stars are due to an ISM bow wave or from a debris disk. By collectively comparing their respective structure and dust grain properties, we can discern whether one of the external mechanisms is more likely.  The fundamental comparison is that ISM grains would give rise to hotter emission compared to debris disk grains at a fixed stellocentric distance. Or equivalently that for an observed temperature the stellocentric distance of small ISM dust grains would be farther out than for larger debris disk dust, as was noticed with the original detection of debris disk hosting stars which ruled out ISM interactions \citep{DB93,ST84}.  The SED degeneracy between temperature and distance resulting from different dust grain sizes and compositions can be broken  
by resolving where the emission is located.

First, we describe the data in Section 2.  Then we determine the spatial extent of the excess emission in Section 3.  In Section 4, we determine the temperature of the dust and the host star properties.  In Section 5, we test the spatial scale and dust properties against bow wave models to break the SED degeneracy and determine which mechanism is more likely.  We also re-affirm the significance of the correlation between IR excesses and Lambda Boo stars in Section 6. In Section 7, we explore what connection debris disks may have to the Lambda Boo-like spectroscopic properties. Finally in Section 8, we attempt to connect observations to the mechanisms at work which could generate the abundance anomaly.

\section[]{\textit{Herschel} Data}

The \textit{Herschel Space Observatory}\footnote{\textit{Herschel} is an ESA space observatory with science instruments provided by European-led Principal Investigator consortia and with important participation from NASA} is a European Space Agency (ESA) Infrared and Sub-millimeter observatory with a monolithic 3.5 meter mirror orbiting at the second Earth-Sun Lagrangian point, or L2 \citep{PG10}. The wavelength range of 55 to 670 $\mu$m is largely opaque from the ground which necessitates space-based observations to find far-IR emission. In total 9 confirmed Lambda Boo stars were imaged with the Photoconductor Array Camera and Spectrometer, or PACS \citep{AP10}.  Two were observed as part of the DEBRIS Open Time Key Programme (HD 110411 ($\rho$ Vir), HD 125162 ($\lambda$ Boo)) and were found to host resolved debris disks by \cite{MB13}. Six of the stars were observed as a targeted PI proposal (HD 11413, HD 30422, HD 31295, HD 183324, HD 198160, HD 221756) to look specifically at Lambda Boo stars with previously known excesses (see Table \ref{obsid}). All eight had broadband imaging at 100 and 160 $\mu$m. Another Lambda Boo star (HD 74873) was observed under another targeted PI proposal and has 70 and 160 $\mu$m imaging (Morales et al. submitted). PACS was designed to take dual band imaging simultaneously in either 100/160 $\mu$m or 70/160 $\mu$m configurations, so no stars have observations at all 3 wavelengths.  HR 8799 was also observed with \textit{Herschel} \citep{BM14B} and is found to host a resolved debris disk with a `mild' Lambda Boo-like abundance anomaly \citep{RG99,SK06}.  Given the varied nature of spectroscopic detections, we choose to limit our sample of stars to those which have been considered confirmed Lambda Boo stars in the literature \citep{GC93},  although we do not find HR 8799 to be inconsistent with the conclusions of this paper. 

\textit{Herschel} PACS images are produced as mini-scan maps with the Herschel Interactive Pipeline Environment \citep[HIPE;][]{OT10}.  The PACS scanning strategy covers the same region of sky multiple times which can be combined in a `drizzle' method to produce an image with better pixel sampling at the cost of correlated noise \citep{FH02}.  This is a beneficial method given that resolving the emission is one of our primary science goals.  At 70/100 $\mu$m, the pixel scale is 1 arsecond per pixel while 160 $\mu$m maps are 2 arcseconds per pixel.

\newcommand{\imagecaption}{70/100 $\mu$m and 160 $\mu$m images of all known Lambda Boo stars targeted by \textit{Herschel} and analysed in this paper.  The field of view (FOV) in each subpanel is $80\arcsec\times80\arcsec$ and each image is individually scaled linearly to the minimum and maximum values. The `drizzle' map is shown on the left and the residual after PSF subtraction on the right with surface brightness (SB) in milli-Janskys per arcsecond squared. White arrows indicate the direction of stellar proper motion. Relative star-cloud motion is estimated in most cases to be $\sim$15$^{\circ}$ from the stellar motion.  Vectors are normalized to a uniform length to indicate direction and not velocity.  There is no preference for excess emission to be in the direction of motion for any of the stars. For HD 74873, HD 183324 and HD 221756, background sources at 70/100 $\mu$m have been removed with a PSF fit from the residual images to highlight the emission associated with the star. At 160 $\mu$m, background sources are likely blending with the excess stellar emission and were pre-subtracted from the `drizzle' maps, based on positions in the 70/100 $\mu$m images, in order to measure excess emission associated with the star.  Note that HD 74873 was the only star observed at 70~$\mu$m instead of 100~$\mu$m \label{psf_100}}

\begin{table*}
\caption{Measured extent of emission around Lambda Boo stars at 70/100 and 160 $\mu$m. The table shows the angular size of emission assuming convolved Gaussians to constrain the spatial scale of far-IR emission.  The radii of the emission ($\rm  R_{outer}$) are the projected sizes given the known distance to each star from \textit{Hipparcos} \citep{FL07}. Note that HD 74873 is the only star observed at 70~$\mu$m, while the rest were observed at 100~$\mu$m. \label{angles_100}}
\begin{tabular}{cccccccc}
\hline
Star & Distance & \multicolumn{3}{c}{70/100 $\mu$m} & \multicolumn{3}{c}{160 $\mu$m}  \\ 
(HD) & d (pc) & $\theta_{\rm measured}$ ($\arcsec$) & $\theta_{\rm sky}$ ($\arcsec$) & $\rm R_{outer}$ (AU) & $\theta_{\rm measured}$ ($\arcsec$) & $\theta_{\rm sky}$ ($\arcsec$) & $\rm R_{outer}$ (AU) \\
\hline
11413 & 77 & $8.08\pm0.12$~~ & $4.22\pm0.23$ & $162\pm9$~~ & $12.75\pm0.17$~~ & ~~$7.00\pm0.31$ & $270\pm12$~~ \\
30422 & 56 & $7.95\pm0.13$~~ & $3.97\pm0.26$ & $111\pm7$~~ & $12.31\pm0.40$~~ & ~~$6.17\pm0.80$ & $173\pm23$~~ \\
31295 & 36 & $9.95\pm0.09$~~ & $7.18\pm0.12$ & $129\pm2$~~ & $15.22\pm0.89$~~ & $10.87\pm1.25$ & $193\pm22$~~ \\
74873 & 54 & $6.22\pm0.16$~~ & $2.98\pm0.33$ & ~~$80\pm9$~~ & $13.05\pm0.89$~~ & ~~$7.49\pm1.57$ & $202\pm42$~~ \\
110411~~ & 36 & $9.75\pm0.07^{\ddag}$ & $7.09\pm0.09$ & $128\pm1$~~ & $12.90\pm0.19^{\ddag}$ & ~~$7.28\pm0.33$ & $131\pm6$~~~~ \\
125162~~ & 30 & $9.31\pm0.10^{\ddag}$ & $6.47\pm0.14$ & ~~$97\pm21$ & $13.62\pm0.39^{\ddag}$ & ~~$8.49\pm0.61$ & $127\pm9$~~~~ \\
$183324^{\dagger}$ & 50 & $7.87\pm0.45$~~ & $3.80\pm0.95$ & $116\pm29$ & $13.84\pm0.61$~~ & $~~8.84\pm0.95$ & $270\pm29^{\dagger}$ \\
198160~~ & 76 & $7.11\pm0.17$~~ & $1.75\pm0.87$ & $< 86^{\star}$ & $11.94\pm0.51$~~ & ~~$5.40\pm1.15$ & $246\pm43$~~ \\
$221756^{\dagger}$ & 80 & $7.05\pm0.42$~~ & $1.49\pm1.44$ & $< 116^{\star}$ & $14.60\pm1.15$~~ & $~~8.12\pm1.69$ & $326\pm68^{\dagger}$ \\
\hline
\end{tabular}\\
\raggedright
\begin{footnotesize}
\hspace{2cm}$^\dagger$ Sources are likely contaminated with nearby emission.\\
\hspace{2cm}$^\star$ Measurements errors were below PSF resolution limit and therefore should be treated as upper limits.\\
\hspace{2cm}$^\ddag$ Adopted from \cite{MB13}.\\
\end{footnotesize}
\end{table*}

\begin{figure*}
\label{psf_100}
\begin{tabular}{ll}
\hspace{-3mm}\subfigure{\includegraphics[width=0.525\textwidth,clip,trim=0.3mm 6mm 8.5mm 0mm]{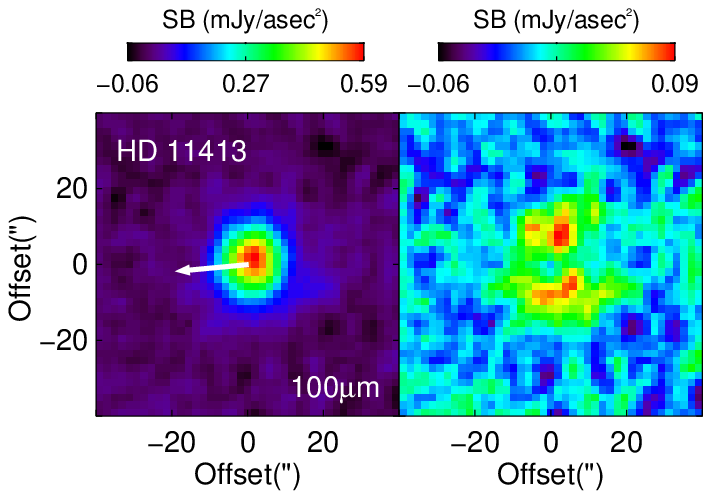}} &
\hspace{-5mm}\subfigure{\includegraphics[width=0.465\textwidth,clip,trim=8.5mm 6mm 8.5mm 0mm]{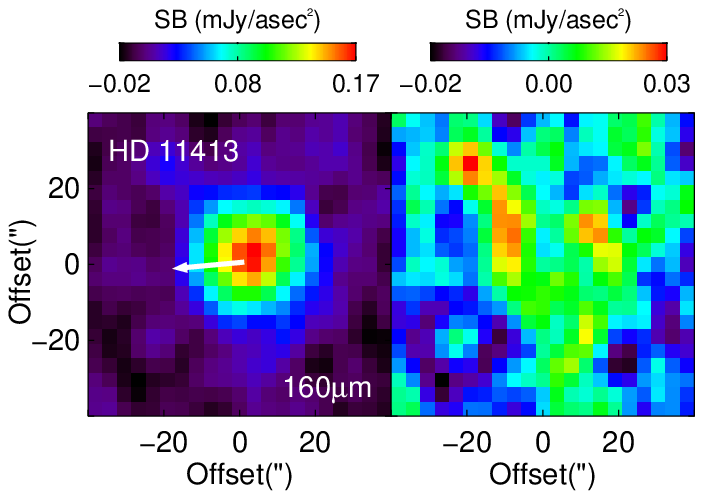}}\\
\end{tabular}
\vspace{-7mm}
\begin{tabular}{ll}
\hspace{-3mm}\subfigure{\includegraphics[width=0.525\textwidth,clip,trim=0.3mm 6mm 8.5mm 0mm]{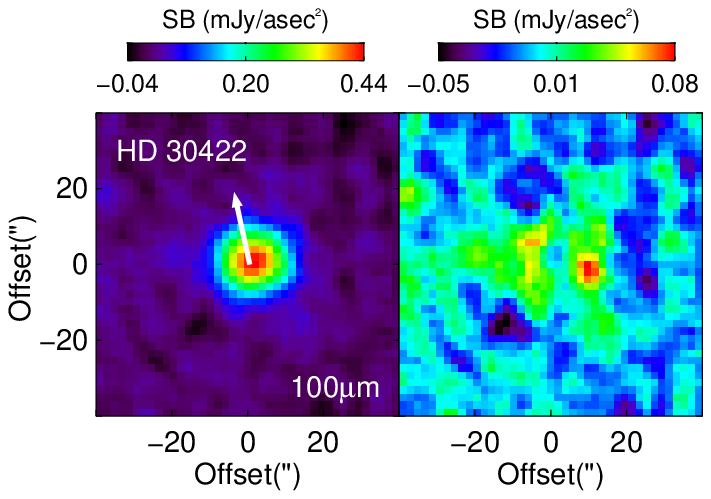}} &
\hspace{-5mm}\subfigure{\includegraphics[width=0.465\textwidth,clip,trim=8.5mm 6mm 8.5mm 0mm]{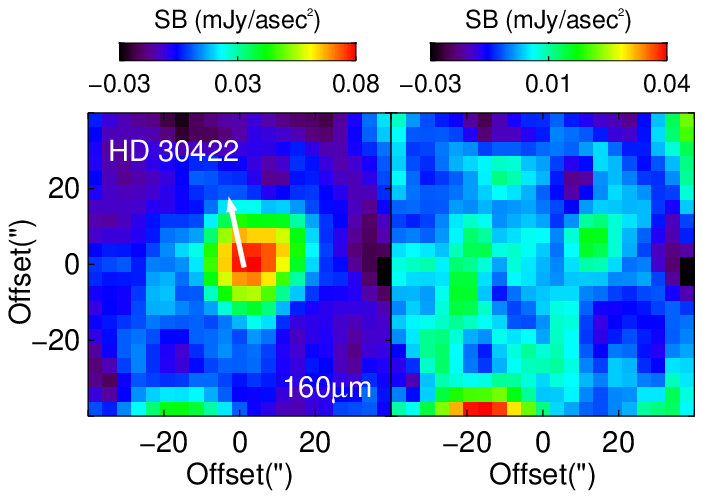}}\\
\end{tabular}
\vspace{-7mm}
\begin{tabular}{ll}
\hspace{-3mm}\subfigure{\includegraphics[width=0.525\textwidth,clip,trim=0.3mm 6mm 8.5mm 0mm]{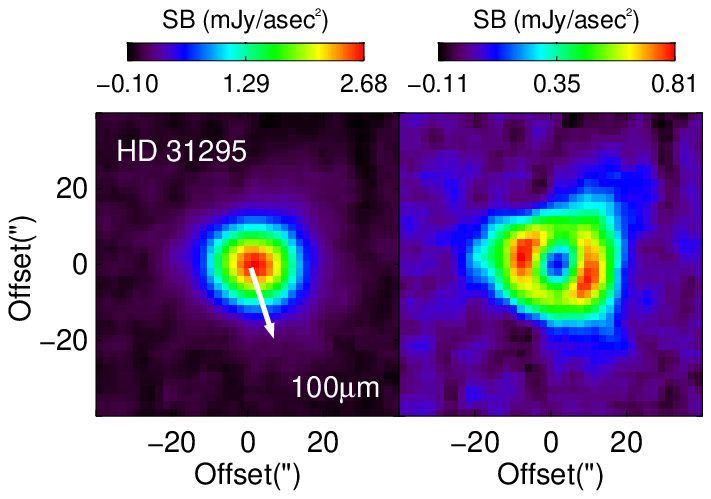}} &
\hspace{-5mm}\subfigure{\includegraphics[width=0.465\textwidth,clip,trim=8.5mm 6mm 8.5mm 0mm]{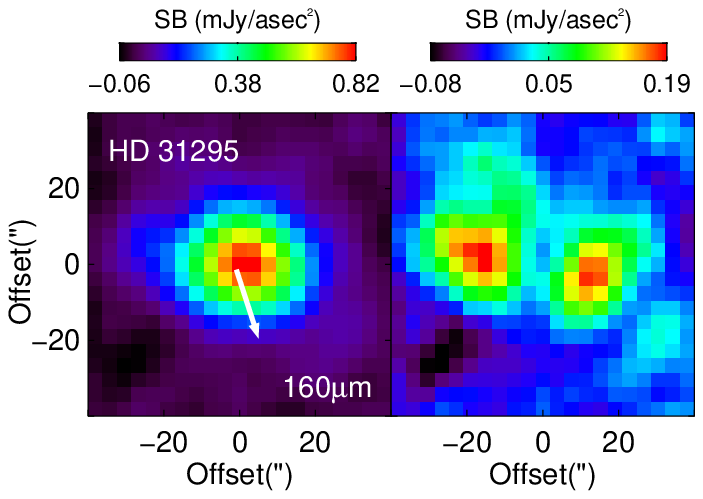}}\\
\end{tabular}
\caption{\imagecaption}
\end{figure*}

\setcounter{figure}{0}
\begin{figure*}
\centering
\begin{tabular}{ll}
\hspace{-3mm}\subfigure{\includegraphics[width=0.525\textwidth,clip,trim=0.3mm 6mm 8.5mm 0mm]{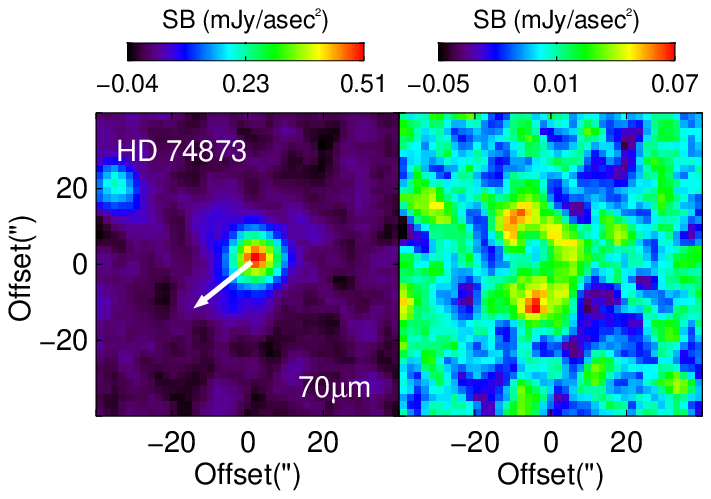} } &
\hspace{-5mm}\subfigure{\includegraphics[width=0.465\textwidth,clip,trim=8.5mm 6mm 8.5mm 0mm]{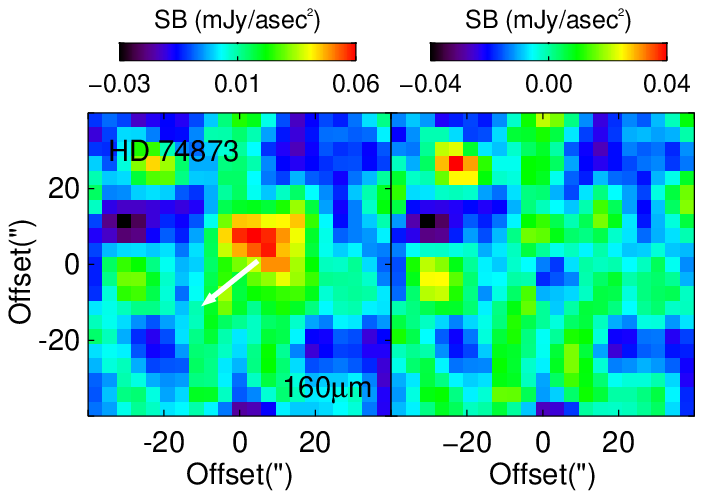}} \\
\end{tabular}
\vspace{-7mm}
\begin{tabular}{ll}
\hspace{-3mm}\subfigure{\includegraphics[width=0.525\textwidth,clip,trim=0.3mm 6mm 8.5mm 0mm]{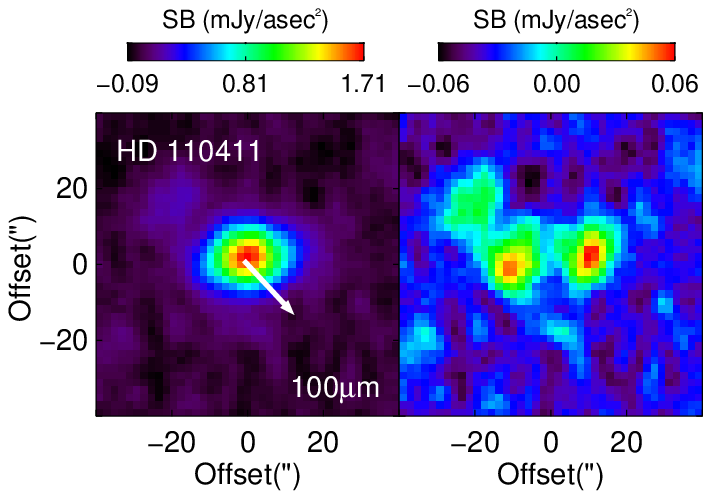}} &
\hspace{-5mm}\subfigure{\includegraphics[width=0.465\textwidth,clip,trim=8.5mm 6mm 8.5mm 0mm]{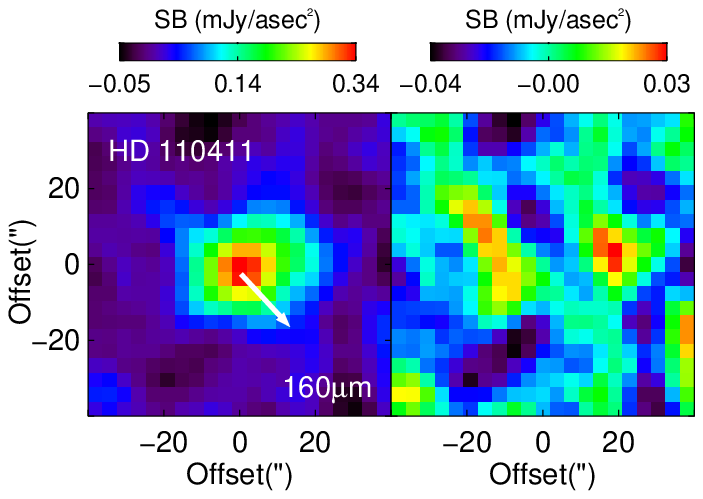}}\\
\end{tabular}
\vspace{-7mm}
\begin{tabular}{ll}
\hspace{-3mm}\subfigure{\includegraphics[width=0.525\textwidth,clip,trim=0.3mm 6mm 8.5mm 0mm]{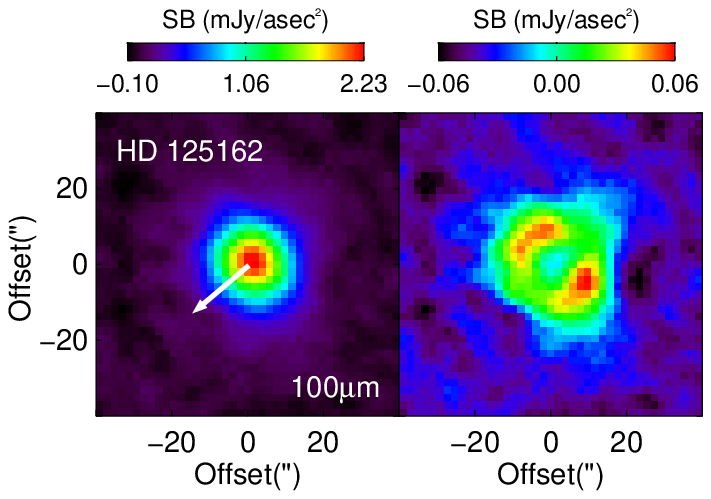}} &
\hspace{-5mm}\subfigure{\includegraphics[width=0.465\textwidth,clip,trim=8.5mm 6mm 8.5mm 0mm]{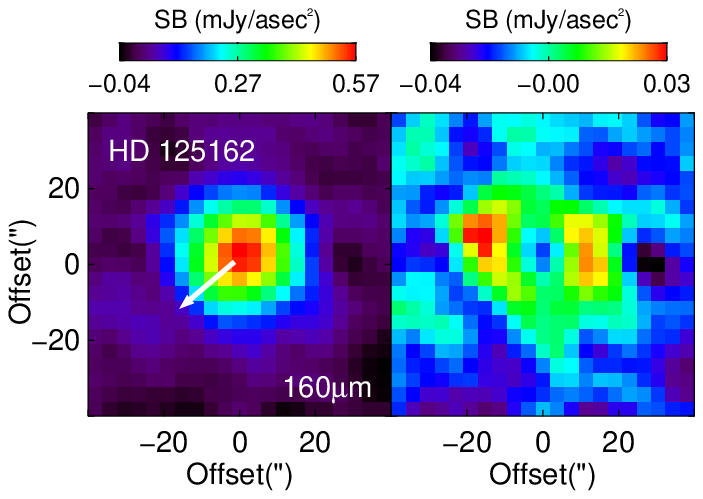}}\\
%\hline
\end{tabular}
\centering
\caption{Continued.}
\vspace{5cm}
\end{figure*}

\setcounter{figure}{0}
\begin{figure*}
\centering
\begin{tabular}{ll}
\hspace{-3mm}\subfigure{\includegraphics[width=0.525\textwidth,clip,trim=0.3mm 6mm 8.5mm 0mm]{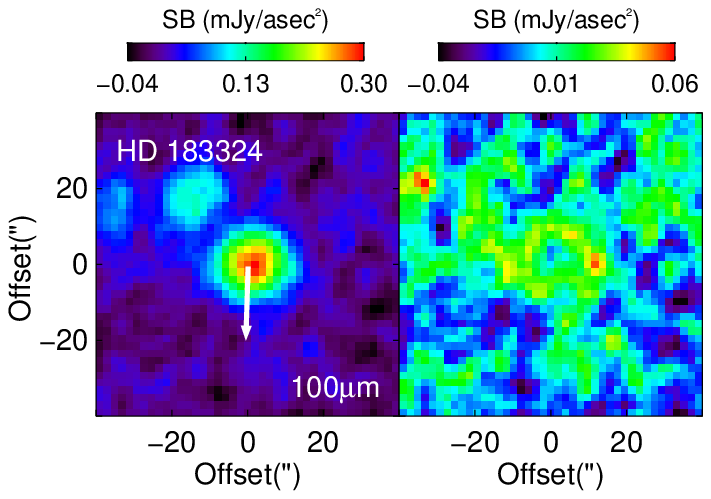}} &
\hspace{-5mm}\subfigure{\includegraphics[width=0.465\textwidth,clip,trim=8.5mm 6mm 8.5mm 0mm]{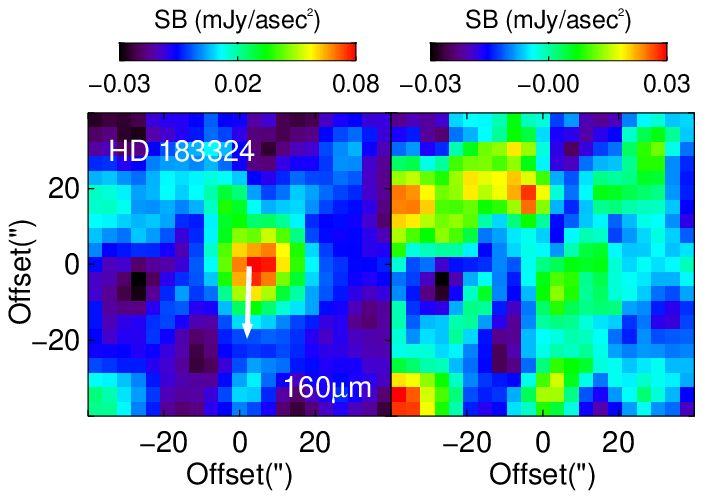}}\\
\vspace{-7mm}
\end{tabular}
\begin{tabular}{ll}
\hspace{-3mm}\subfigure{\includegraphics[width=0.525\textwidth,clip,trim=0.3mm 6mm 8.5mm 0mm]{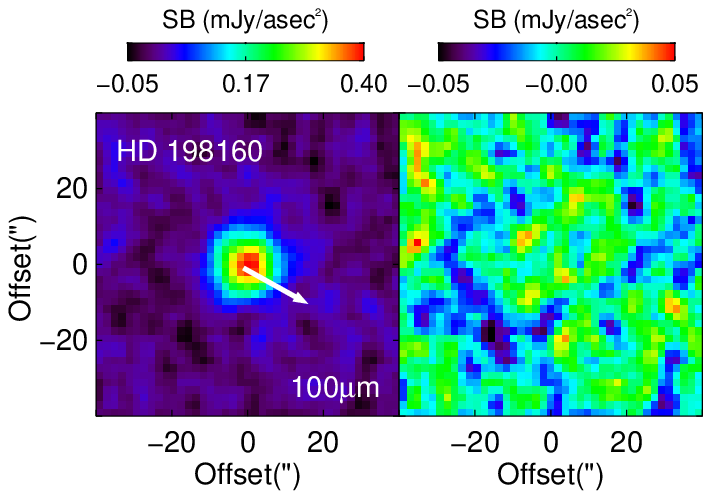}} &
\hspace{-5mm}\subfigure{\includegraphics[width=0.465\textwidth,clip,trim=8.5mm 6mm 8.5mm 0mm]{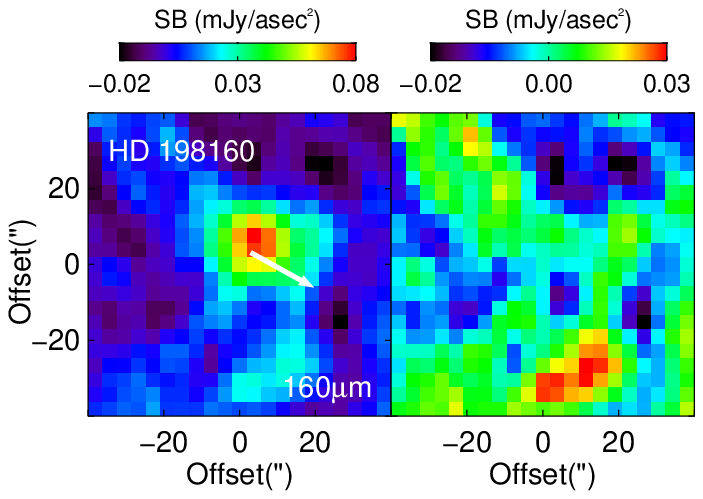}}\\
\end{tabular}
\vspace{-7mm}
\begin{tabular}{ll}
\hspace{-3mm}\subfigure{\includegraphics[width=0.525\textwidth,clip,trim=0.3mm 6mm 8.5mm 0mm]{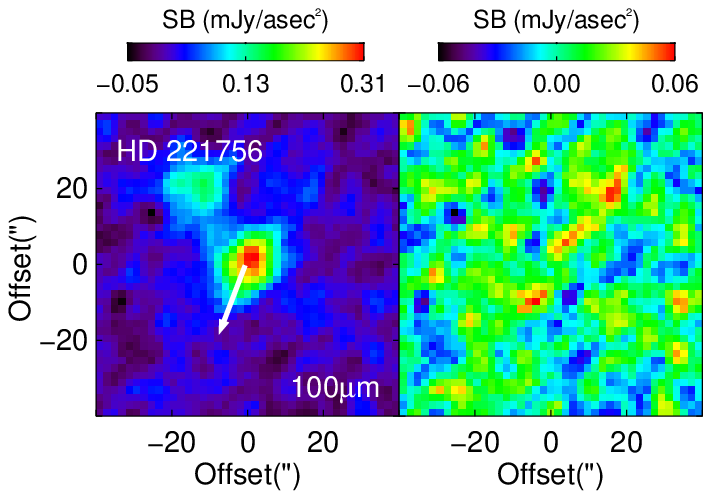}} &
\hspace{-5mm}\subfigure{\includegraphics[width=0.465\textwidth,clip,trim=8.5mm 6mm 8.5mm 0mm]{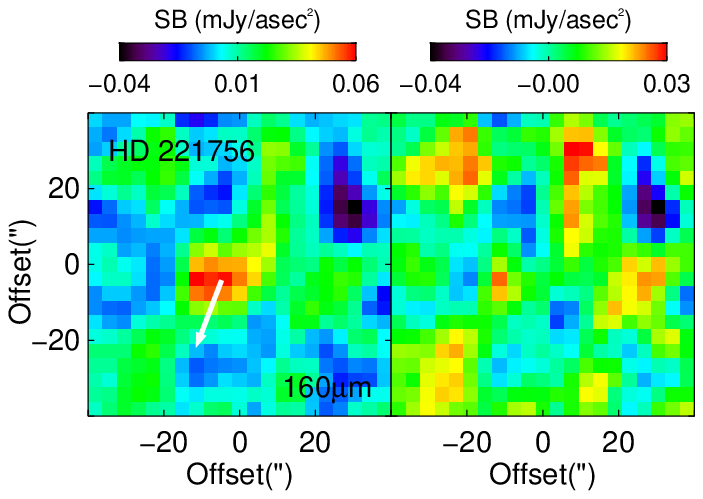}}\\
\end{tabular}
\centering
\caption{Continued.}
\vspace{5cm}
\end{figure*}
\renewcommand{\thefigure}{\arabic{figure}}

\begin{figure*}
\centering
\begin{tabular}{c}
\subfigure{\includegraphics[width=0.75\textwidth,clip,trim=0.0mm 1.5mm 0mm 0mm]{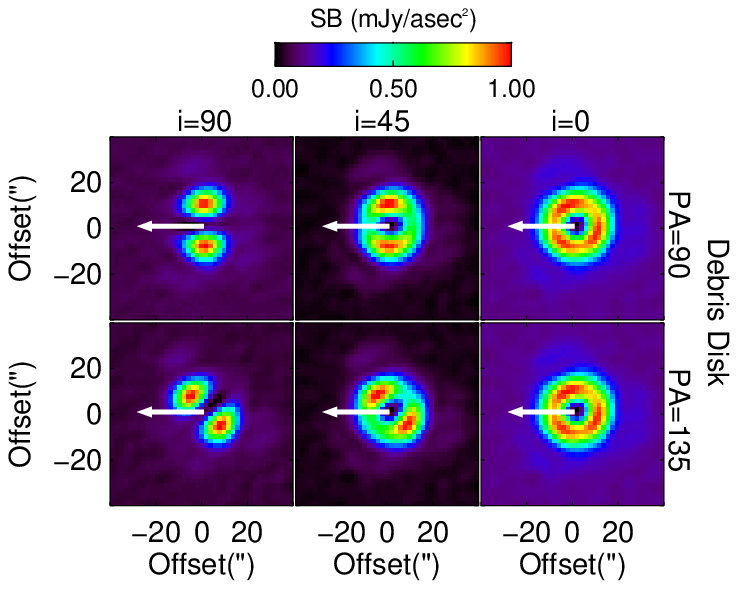}} \\

\subfigure{\includegraphics[width=0.75\textwidth,clip,trim=0mm 1.5mm 0mm 0mm]{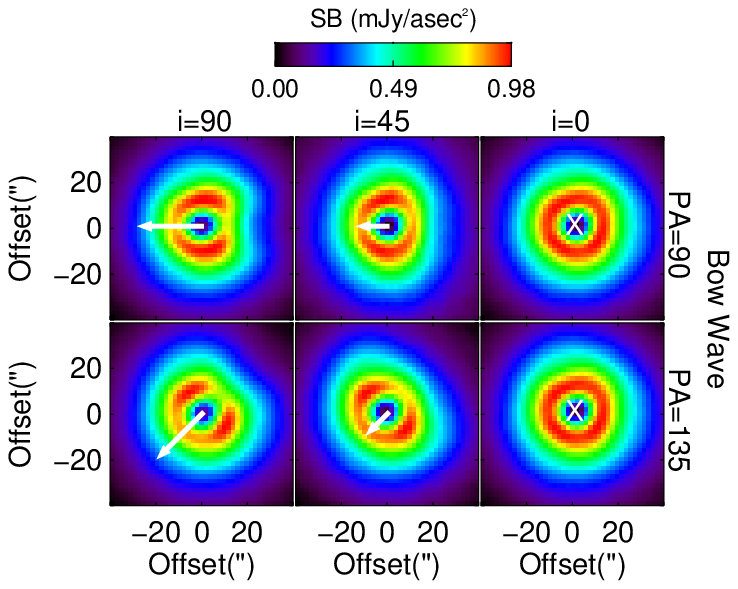}}\\

\end{tabular}
\centering
\caption{The top set of figures show a single debris disk model at different viewing angles as it would be viewed with \textit{Herschel} PACS at 100 $\mu$m.  The inclinations are the angle between the proper motion vector and the line of sight.  While the disk is inclined by the angle between its orthogonal axis relative to the line of sight.  The PA is the angular position on sky.  The disk model is a 80 to 120 AU ring around a nominal A-type star at 30 pc on top. The bottom set of figures is a bow wave model imaged from the same viewing angles, again around an A-type star at 30 pc, based on the models described in  \protect\cite{GA08}.  The models assume an inner avoidance radius $\rm r_{ave}$ of 5 AU and a $\rm R_{outer}$ of the surrounding spherical cloud to be 905 AU, as was typical with models from \protect\cite{JM09} using \textit{Spitzer} data fit to Lambda Boo stars IR excesses. The models are convolved with a PACS beam and PSF subtracted by peak scaling to compare with the morphological structure of the residual excess emission seen in Figure \ref{psf_100}.  By definition, the bow wave models' bilateral symmetry is perpendicular to the proper motion seen by the white arrows.  Debris disks bilateral symmetry on the other hand are not causally connected to their proper motion. Furthermore, the scale of emission in the bow wave models are much more extended due to the ISM cloud around it, which is ultimately necessary to match total flux at longer wavelengths \protect\citep{JM09}. The density of the ISM cloud has been scaled such that it roughly matches the peak flux density of the observations in Figure \ref{psf_100}. \label{PACS_models}}
\vspace{5cm}
\end{figure*}
\renewcommand{\thefigure}{\arabic{figure}}

\section{Extent of Far-IR Excess Emission}
\label{extentemission}

By resolving the spatial scale of the emission, the SED degeneracy between dust grain size and stellocentric radius for a fixed temperature can be broken.  This is done through a Gaussian deconvolution to get the projected size of the emission on the sky. This gives the resolved outer radius of the emission to compare with the dust temperature.  A debris disk model is also favoured if the resolved emission appears compact and bilaterally symmetric through PSF subtraction. 

All images were rotated for north up orientation, east to the left.  Each target star is selected from the centre of the image with a box of 20$\times$20 pixels, in either band, by fitting a two dimensional Gaussian profile with MPFIT2DPEAK in IDL.  These sources were visually identifiable as where the stellar emission was expected to be.  Observations of the calibration star $\gamma$ Draconis were reduced and used in conjunction with these observations to serve as a PSF reference star.  The PSF reference star can be scaled to the peak emission and subtracted from the data images to test for coherent structure indicating emission with a spatial scale larger than the PSF.  All of the sources appear as centrally peaked in reduced images, yet some are plainly resolved with PSF subtraction (see Figure \ref{psf_100}). However, we will not necessarily resolve a bow wave or debris disk (see Section \ref{bowmodels}). The emission around HD 31295, HD 110411, and HD 125162 are well resolved, exhibiting a symmetric structure in the PSF-subtracted emission on either side of the star. HD 11413, HD 30422 and HD 74873 also have faint residual structures which can be seen to either side of the star.  

The morphology of the data in Figure \ref{psf_100} can be qualitatively compared to bow wave models and debris disk models seen in Figure \ref{PACS_models} which have been convolved with a PACS beam and PSF subtracted in the same manner. Figure \ref{PACS_models} (Top) shows a debris disk from 80 to 120 AU projected at a distance of 30 parsecs around an A-type star at multiple inclinations and position angles. A distance of 30 pc is consistent with the best resolved stars in our sample.  The bow wave models of Figure \ref{PACS_models} (Bottom) represent a spherical ISM cloud with an inner avoidance radius of 5 AU and an outer radius of 905 AU assuming the dust interaction geometry from \cite{GA08}, which creates a cavity within the cloud. An inner avoidance radius of 5 AU represents the scenario with the least possibility of resolving the bow wave in our sample (see section \ref{bowmodels}). The uniform density of the cloud has been arbitrarily scaled to have a resulting peak surface brightness of $\sim$1 mJy/(asec)$^2$, typical of the data. \textit{Herschel's} pointing accuracy is not precise enough to astrometrically measure the location of the star and/or use optical imaging to determine if the excess emission were offset from the star.  As a result, PSF centering with a Gaussian could cause asymmetric emission of a bow wave to appear more symmetric.   

In Figure \ref{PACS_models}, the large scale emission from the ISM cloud is readily apparent.  It also has a ``horseshoe" shape as the bow wave's influence on the cloud is only partially resolved. On the other hand, debris disk emission is relatively more compact to reach the same peak surface brightness of $\sim$1 mJy/(asec)$^2$.  The debris disk models also have a double peak symmetry except when face-on.  Furthermore, the symmetry in the bow wave models is causally connected to the proper motion, while the debris disk emission symmetry is independent of proper motion. Given a maximal offset of $90^{\circ}$, the difference between the stellar proper motion and the relative proper motion for a star travelling 25 km/s with an ISM cloud moving 7 km/s is a deviation of $\sim$15$^{\circ}$. It should also be noted that there is a third velocity into and out of the page which is not represented by the white arrows but changes the residual morphology as a function of inclination.  Overall, we find the morphology of the emission to be more consistent with debris disks.

In addition to subtracting a peak-scaled PSF, the outer radial extent of excess emission can be measured in \textit{Herschel} PACS observations by measuring the observed Gaussian FWHM with the expected PSF FWHM \citep{GK12}. The largest FWHM is then used as a measurement of the outer radius of the emission ($\theta_{measured}$).  The true on sky extent of the emission ($\theta_{sky}$) is calculated with simple Gaussian deconvolution with the \textit{Herschel} PACS PSF, as shown in Equation (\ref{decon}). 

\begin{equation}
\label{decon}
\theta_{\rm sky} = \sqrt{\theta_{\rm measured}^{2} - PSF_{\rm fwhm}^{2}}
\end{equation}

\noindent This is repeated for both 70, 100, and 160~$\mu$m where the FWHM of the PSF is 5.46, 6.69, and 10.65 arcseconds, respectively given a 20 ($\arcsec$)/s scan speed \citep{AP10}. The result of this analysis can be seen in Table \ref{angles_100}, where the $\rm R_{outer}$ is the outer limit on the radial extent of the emission in astronomical units (AU) projected at the distance of the star.

In order to estimate the error in the outer radial extent, the maps are randomly measured with apertures of the same size to determine the standard deviation of the noise in the background.  This error is then propagated to the estimated Gaussian fit parameters in MPFIT down to the deconvolution which determines the error in projected AU from the star.  The error in the \textit{Hipparcos} distance measurement is considered negligible in the calculation of the projected radius.  Some error measurements result in FWHM which indicate they could be less than the instrumental FWHM for a point source, which is non-physical, but suggests that the emission is not resolved and can only be used to place upper limits on the outer radial extent of emission.

HD 74873, HD 110411, HD 183324 and HD 221756 are seen to have adjacent, but well separated point sources.  They are perhaps high redshift galaxies which have been found in greater abundance with \textit{Herschel} than predicted \citep{DD14}.  There are no previously identified far-IR galaxies at the observed depths on these patches of sky.  The bright A star nearby also makes optical observations impractical to determine if it is a galaxy without high contrast or star subtraction techniques. Verifying they are galaxies would require further characterization which is unavailable at this time.  Still, the emission is well separated at 70/100 $\mu$m and therefore unlikely associated with the star.  At 160 $\mu$m, poorer resolution causes the sources to blend.  In these cases, the nearby excesses were fit with PSFs and subtracted away before point-source PSF subtraction and angular size measurements were made by using 70/100 $\mu$m data as a positional reference (see Figure \ref{psf_100}).

\begin{table}
\centering
\caption{Photometry of the excess emission around the targeted stars at 70/100 $\mu$m and 160 $\mu$m. Note that HD 74873 is the only star observed at 70 $\mu$m, while the rest were observed at 100 $\mu$m. \label{tab:fluxes_100} }
\begin{tabular}{ccc}
\hline
Star (HD) & 70/100 $\mu$m (mJy) & 160 $\mu$m (mJy) \\
\hline
11413 & $55.8\pm2.6$ & ~~$40.6\pm2.3$ \\
30422 & $40.2\pm3.8$ & ~~$16.4\pm1.5$ \\ 
31295 & $392\pm14$ & $190.7\pm8.4$ \\
~~74873$^\dagger$ & $29.6\pm1.1$ & ~~$13.7\pm2.4$ \\
110411$^\dagger$ & ~~~$154\pm7.0^\ddag$ & ~~~~~$67.3\pm7.0^\ddag$ \\
125162~~ & ~~$272\pm15^\ddag$ & ~~~~$142\pm12^\ddag$ \\
183324$^\dagger$ & $25.4\pm1.1$ & ~~$17.2\pm3.3$ \\
198160~~ & $30.7\pm1.2$ & ~~$14.3\pm1.7$ \\
221756$^\dagger$ & $24.1\pm1.1$ & ~~$12.5\pm2.0$ \\ 
\hline
\end{tabular}\\
\raggedright
\begin{footnotesize}
\hspace{0cm}$^\dagger$ Image required PSF fitting to remove nearby background source.\\
\hspace{0cm}$^\ddag$ Adopted from \cite{MB13}.\\
\end{footnotesize}
\end{table}

\section{Spectral Energy Distributions}
\label{sedsec}

\begin{figure*}
\begin{tabular}{ccc}
\hspace{-3mm}\subfigure{\includegraphics[width=0.33\textwidth]{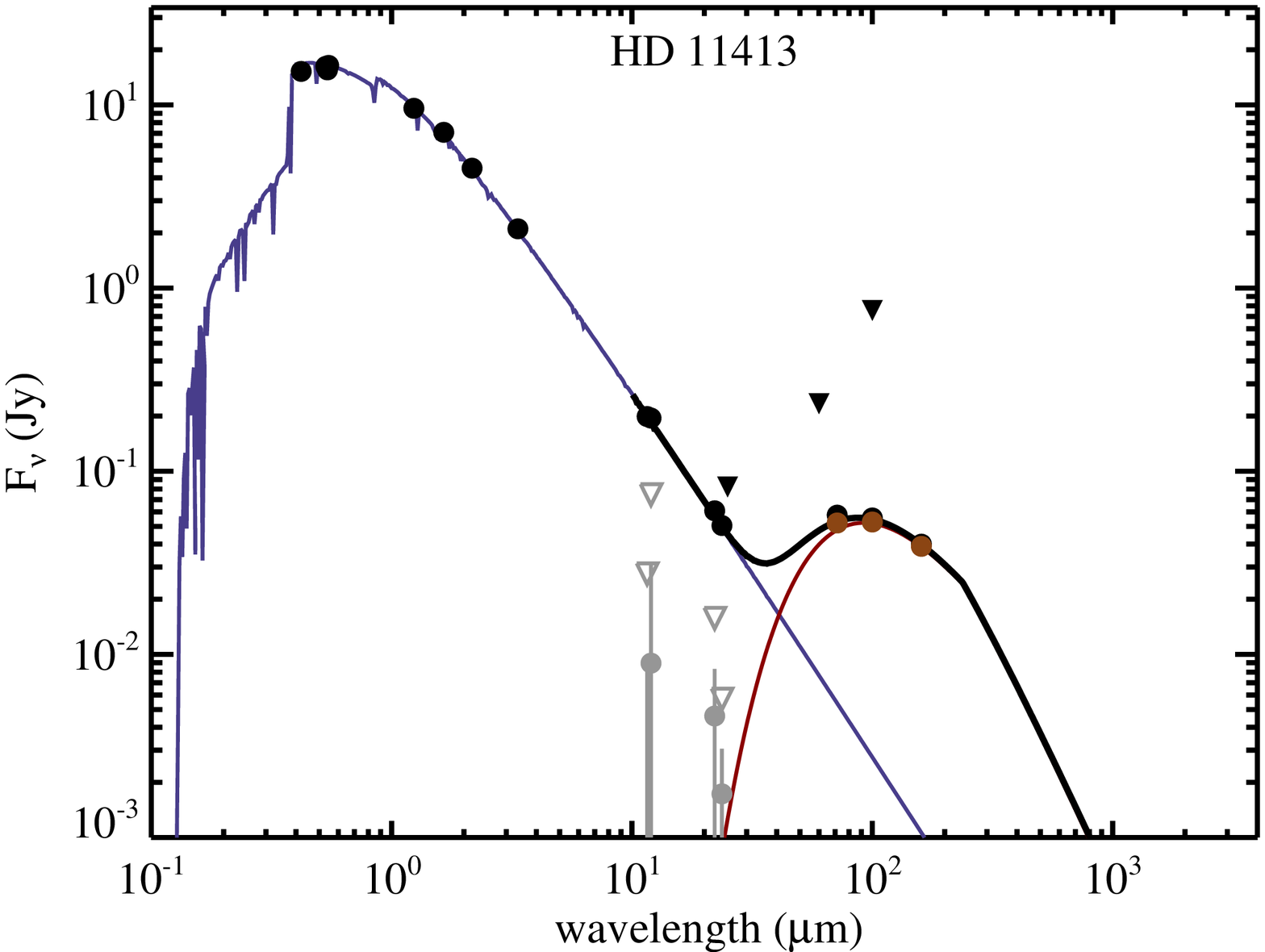}} & \hspace{-4mm}\subfigure{\includegraphics[width=0.33\textwidth]{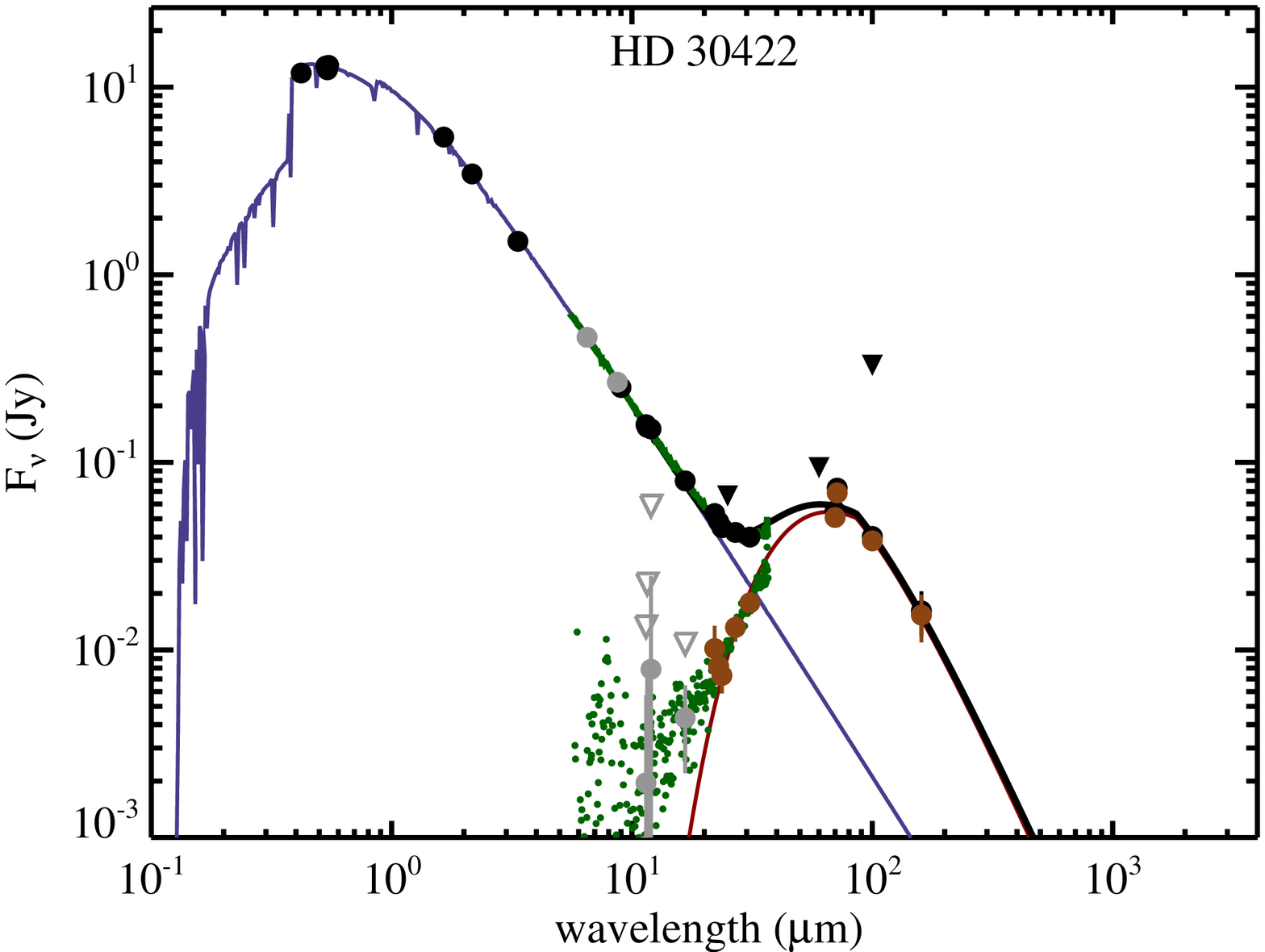}} & \hspace{-4mm}\subfigure{\includegraphics[width=0.33\textwidth]{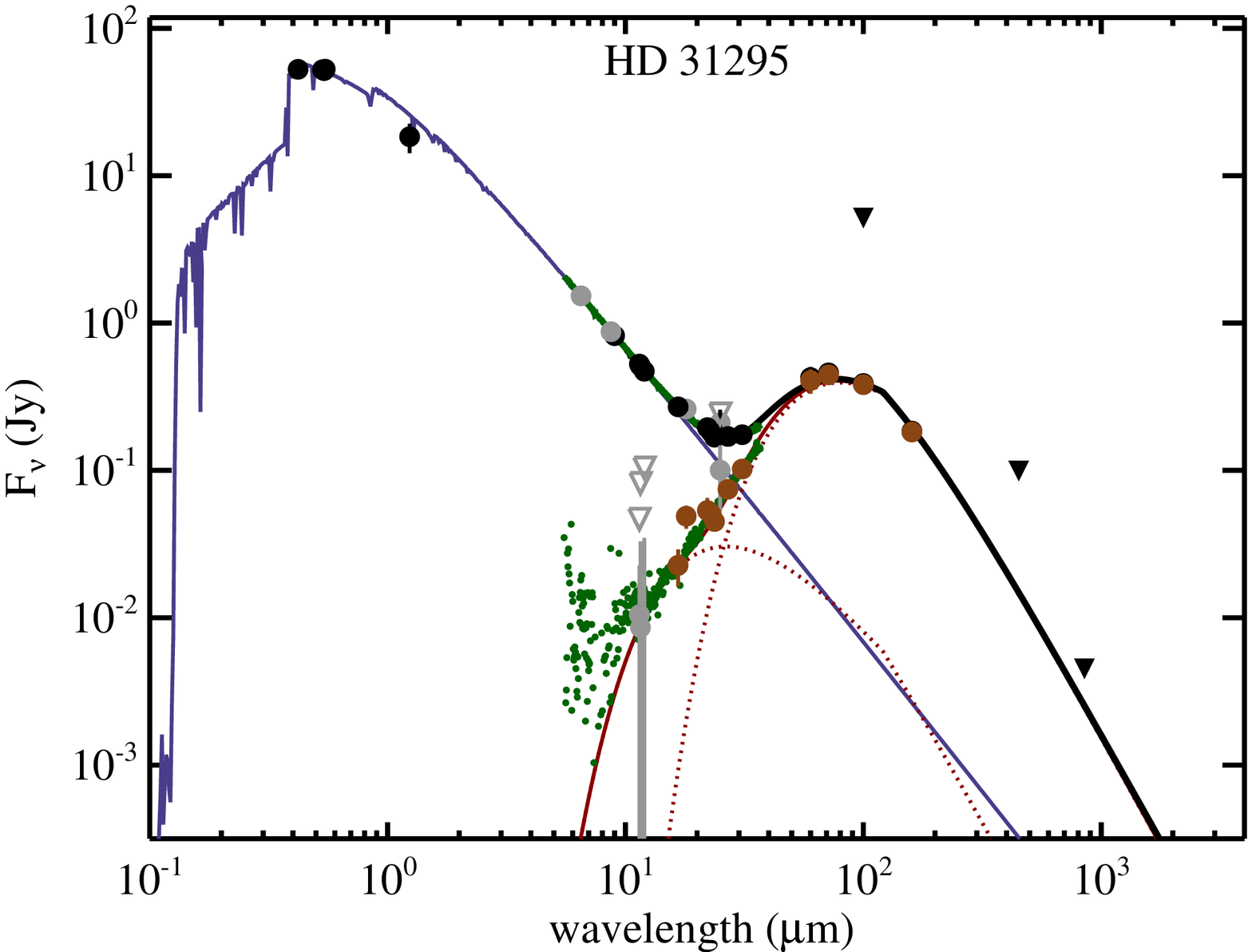}}\\
\hspace{-3mm}\subfigure{\includegraphics[width=0.33\textwidth]{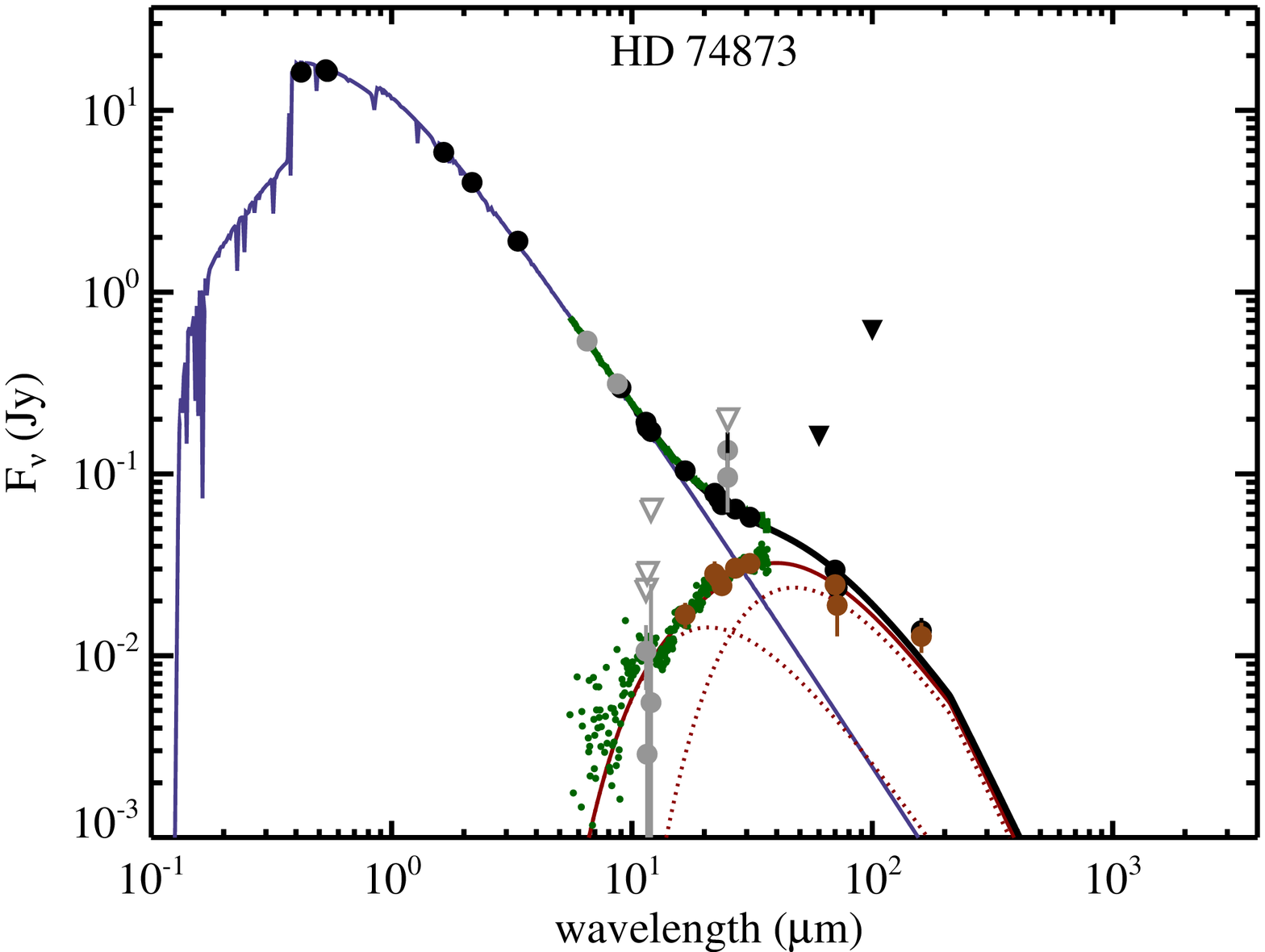}} & \hspace{-4mm}\subfigure{\includegraphics[width=0.33\textwidth]{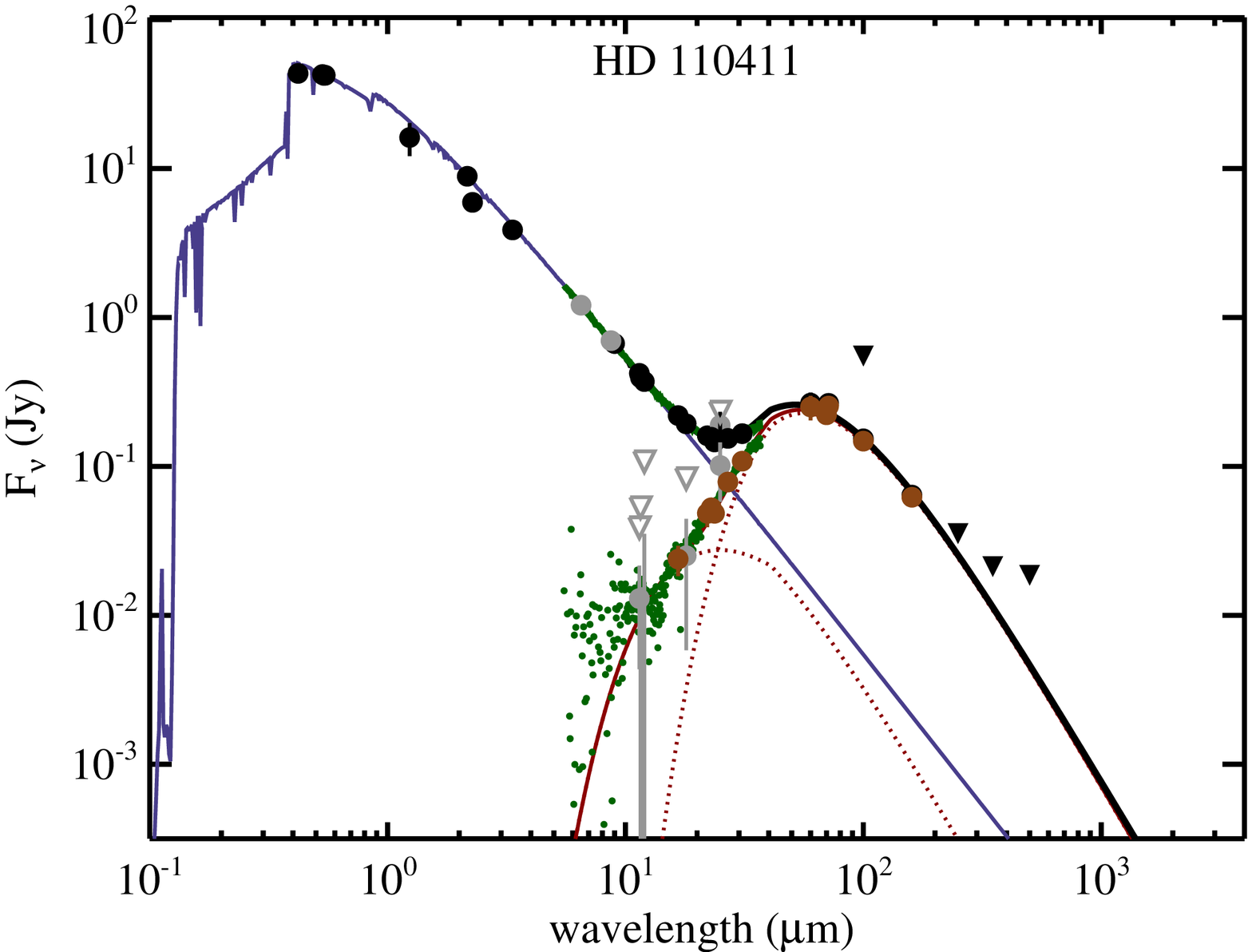}} & 
\hspace{-4mm}\subfigure{\includegraphics[width=0.33\textwidth]{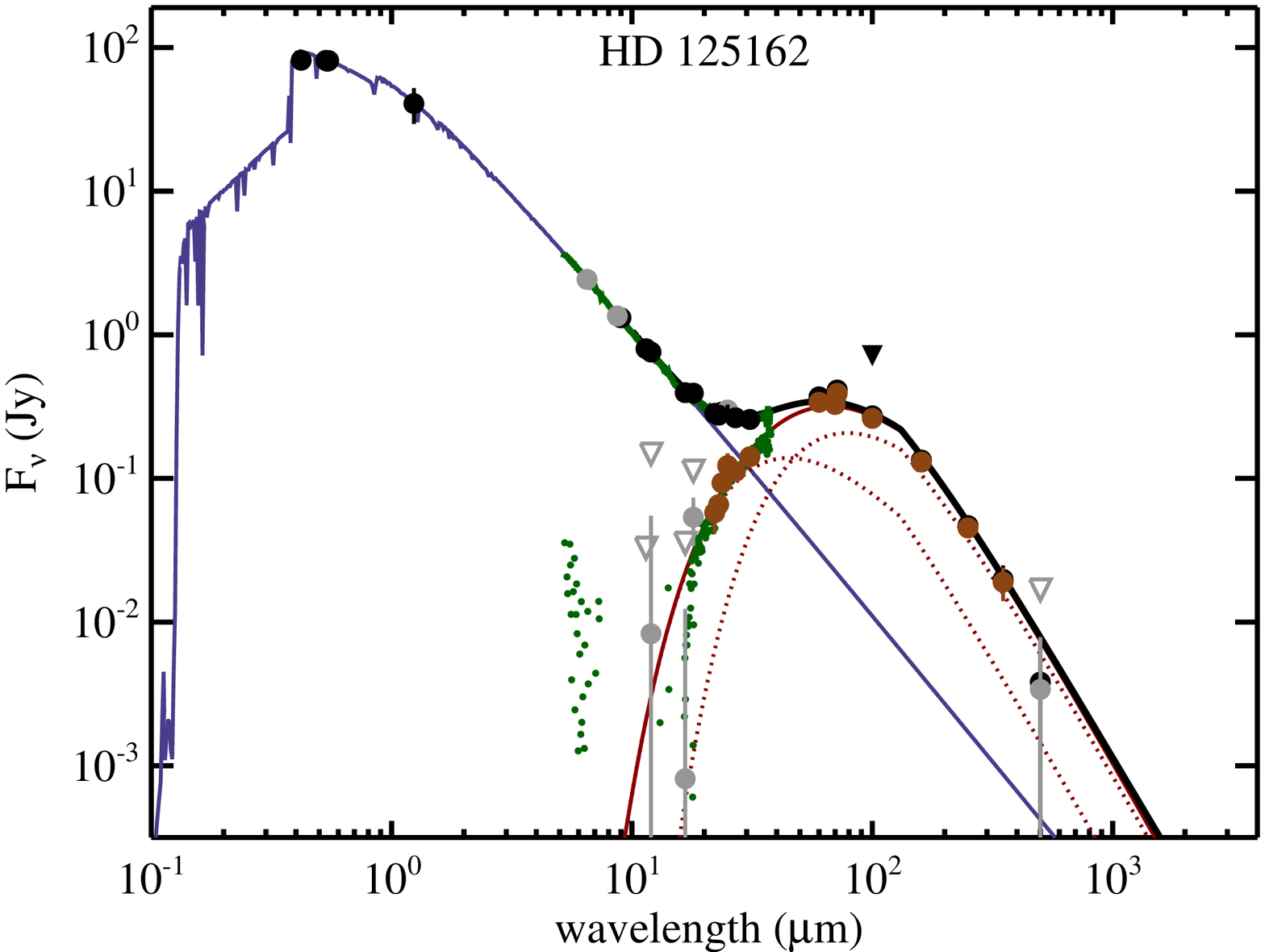}}\\
\hspace{-3mm}\subfigure{\includegraphics[width=0.33\textwidth]{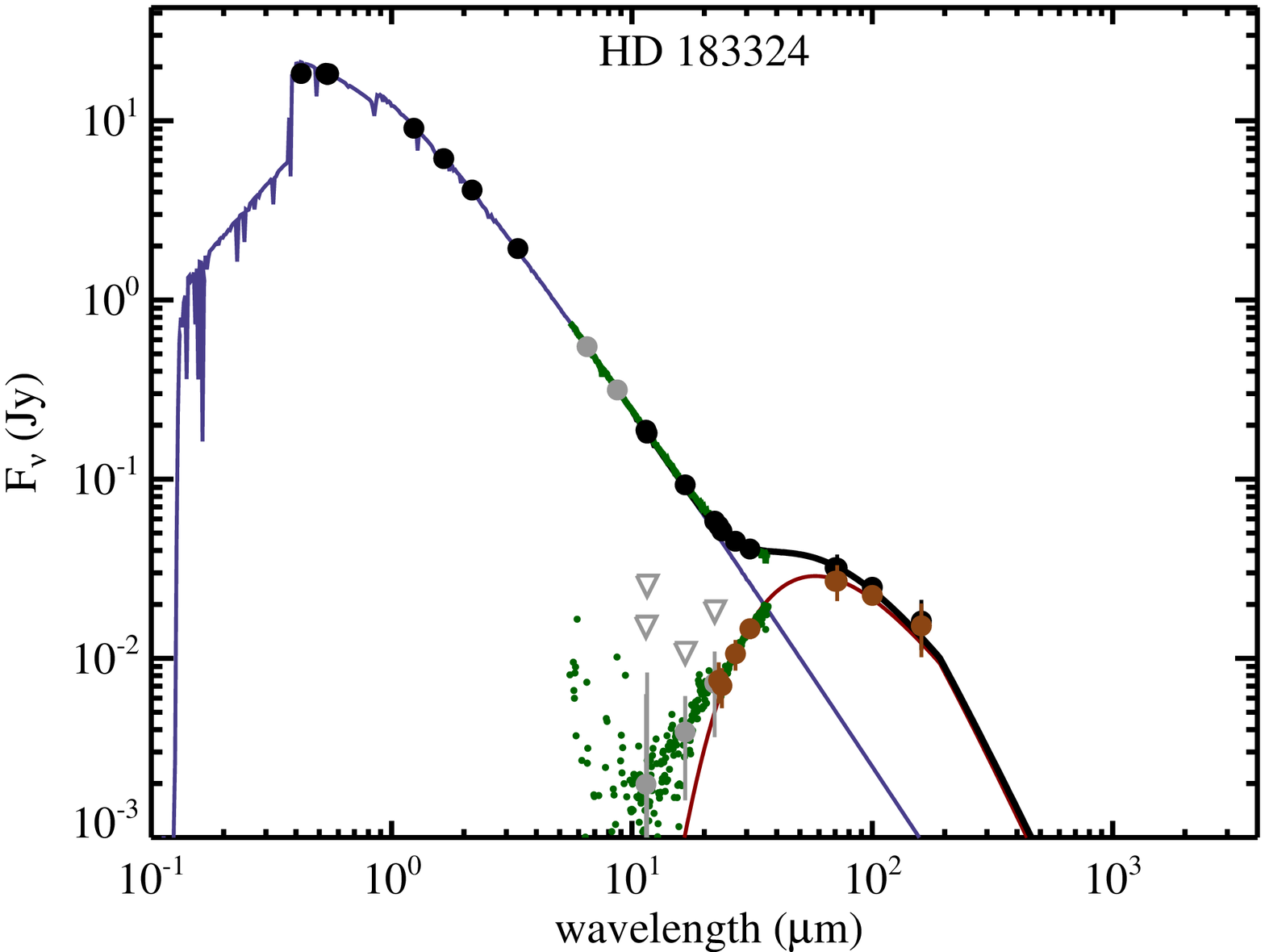}} & \hspace{-4mm}\subfigure{\includegraphics[width=0.33\textwidth]{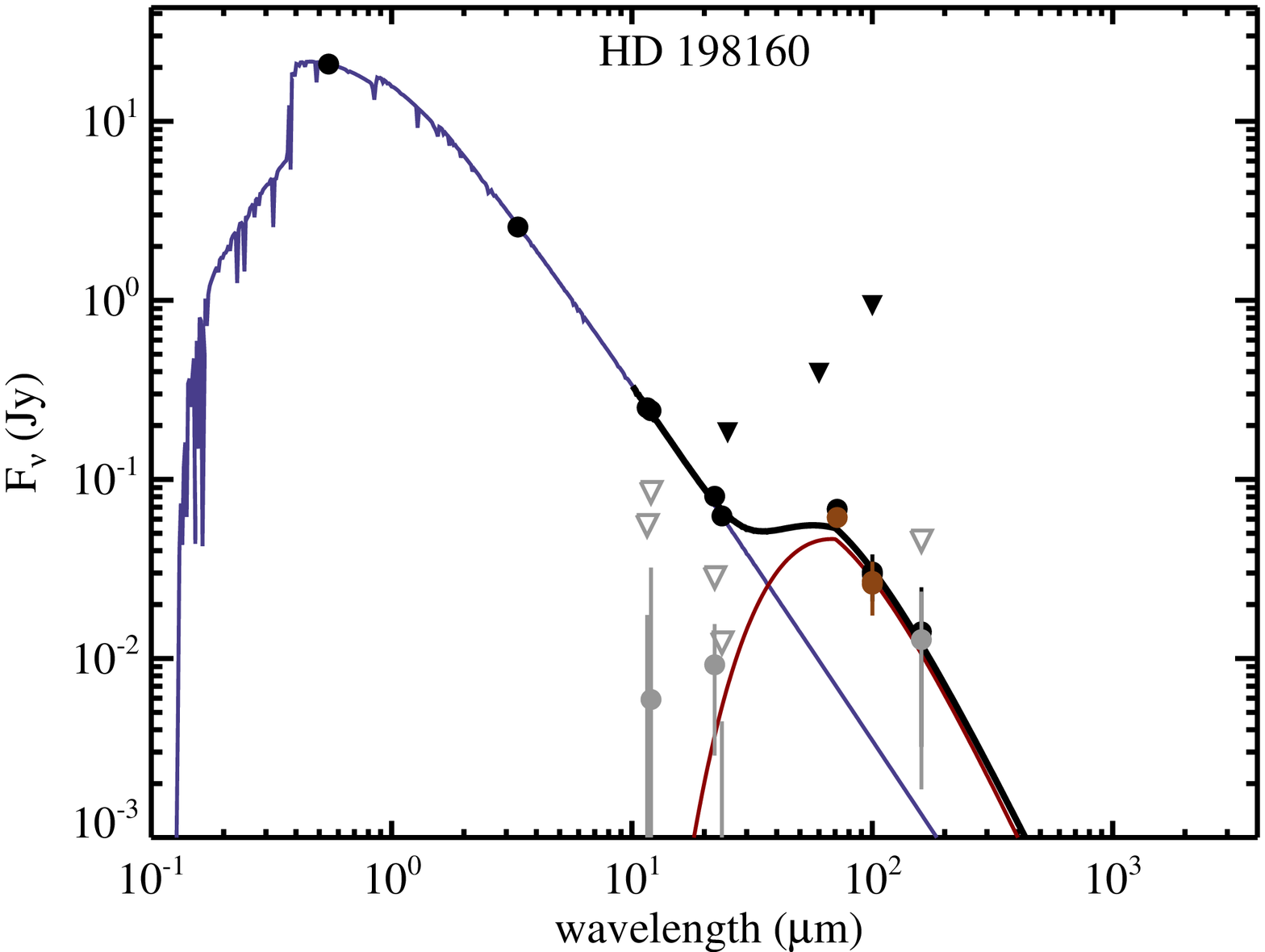}} & 
\hspace{-4mm}\subfigure{\includegraphics[width=0.33\textwidth]{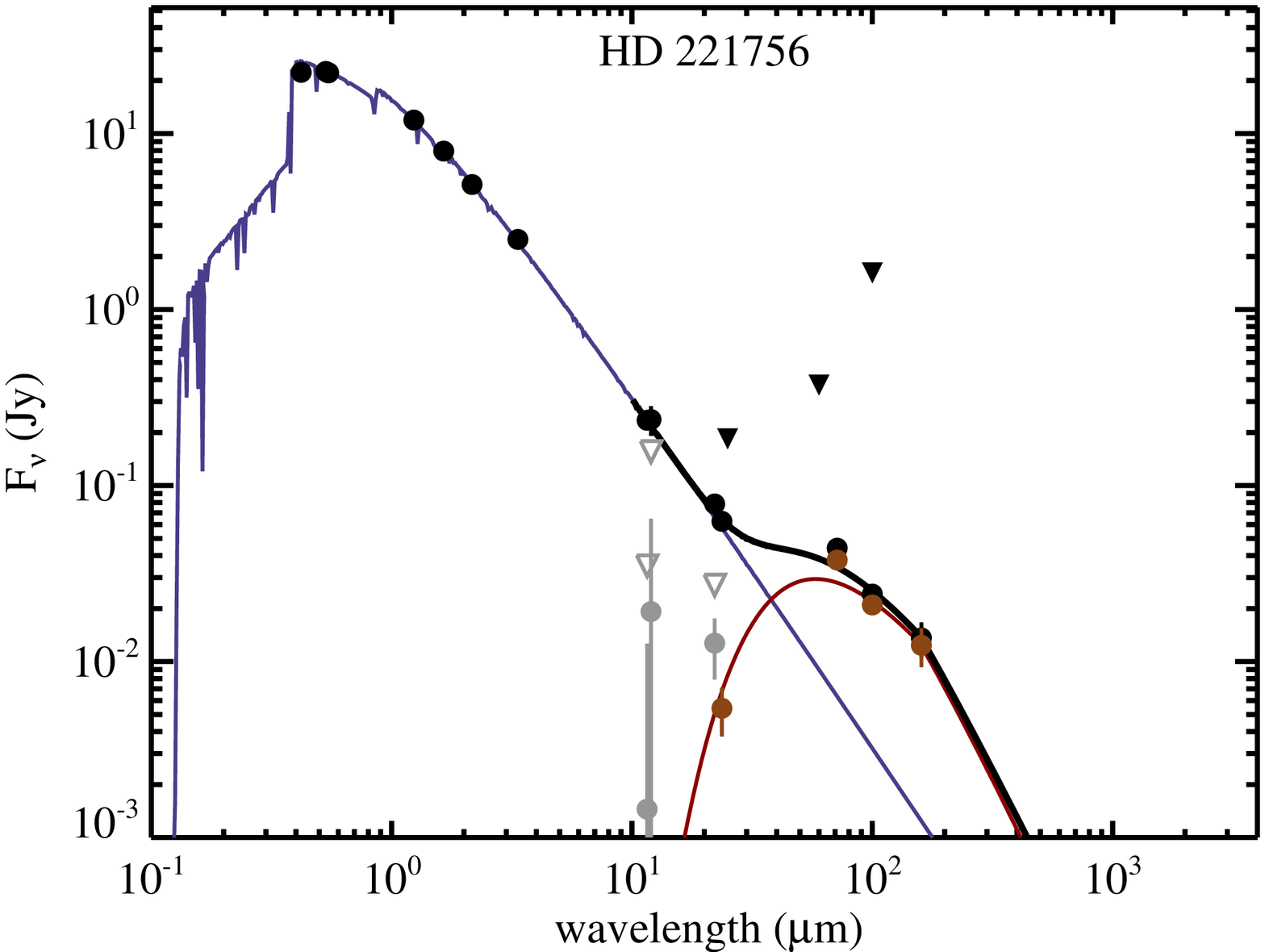}}\\ 
\end{tabular}
\caption{SEDs of all 9 Lambda Boo stars observed by \textit{Herschel} in combination with archival photometry. Measured fluxes are black dots. Black triangles are upper limits from IRAS \citep{MM90} and JCMT \citep{OP13}. Brown dots are excess-only, star subtracted fluxes. Grey dots show star-subtracted fluxes that are consistent with zero at 3$\sigma$, and grey triangles show their associated 3$\sigma$ upper limits. \textit{Spitzer} IRS spectra are shown as green dots. The blue line is a PHOENIX stellar spectrum model fit to optical and near-IR data \citep{BH05}.  Red-brown lines are the blackbody SED fits of excess emission. In some cases, two black body fits are necessary to fit the excess emission and are shown separately as dashed red-brown lines.  The black line is the star+excess SED. \label{sed}}
\label{seds}
\end{figure*}

We now require knowledge of the temperature of the excess emission in order to compare with the spatial extent and break the SED degeneracy due to grain size and composition.

In the 70/100~$\mu$m images the sources were sufficiently separated from contaminating sources that aperture photometry could be used to extract the stellar and excess emission associated with the star.  An aperture radius of 18$\arcsec$ and 36$\arcsec$ for 70/100~$\mu$m and 160~$\mu$m images, respectively, was used for all target stars.  Aperture correction was applied to the fluxes in all bands as described in \cite{ZB14} and can be seen in Table \ref{tab:fluxes_100}. For HD 74873, HD 110411, HD 183324 and HD 221756, there were additional point sources with extended excess emission present.  In those cases, background PSF subtracted images were necessary to decorrelate nearby emission from emission associated with the star itself before measuring the flux with an aperture. Again, it can be seen that the adjacent point sources of emission are distinctly separated at 70/100~$\mu$m and only appear connected via overlapping wings of the PSFs at 160~$\mu$m (see Figure \ref{psf_100}).

The \textit{Herschel} measurements were used in conjunction with archival photometry to construct a multi-wavelength SED from the UV atmospheric cutoff to sub-mm.  When available, photometry from optical surveys \citep{BH98,JM06}, 2MASS \citep{RC03}, \textit{WISE} \citep{EW10}, \textit{Spitzer} \citep{KS06,CC09,FM09}, and \textit{Akari} \citep{DI10} were used. In each case the stellar photosphere was fit with a PHOENIX Ames-COND model \citep{BH05}, using only the photometry at wavelengths shorter than 10~$\mu$m. The stellar model SED fits determine the $\rm T_{eff}$, and in combination with a known distance, the luminosity seen in Table \ref{tab:stars}. Furthermore, an approximate stellar mass can be tabulated following relationships of modelled main sequence stars taken from \cite{SK82} and \cite{PM13}.

A modified black body spectrum was then fit to the mid- to far-IR observations to model the contribution due to excess dust emission (See Figure \ref{sed}).  The black body equation in Jy/sr can be seen in Equation \ref{bb1} where $h$ is Planck's constant, $c$ is the speed of light, and $k$ is Boltzmann's constant.  The main independent variables over a given wavelength range are the temperature and effective emitting area of dust in steradians ($\Omega$) which can vary to fit the flux density ($\rm F_{\nu}$) of the observations in Jy (Eq. \ref{bb}). Uncertainties are determined from the diagonal elements of the covariance matrices from the least-square fit.  The fit itself is weighted by the measured magnitude uncertainties. 

\begin{equation}
\rm{B(\lambda,T)} = \frac{2hc^{2}}{\lambda^{5}} \frac{1}{e^{\frac{hc}{k \lambda T}}-1}
\label{bb1}
\end{equation}

\begin{equation}
\rm{F_{\nu}} = \int B(\lambda,T) d\Omega
\label{bb}
\end{equation}

Dust grains are inefficient emitters at wavelengths much greater than the grain size and require a modified Rayleigh Jeans tail power law to match observations at wavelengths greater than $\sim$50~$\mu$m.  Therefore, $\beta$ and $\lambda_{0}$ parameterize the slope and wavelength where the modified blackbody intensity deviates from a normal Rayleigh-Jeans tail (see Equation \ref{mbb}).  In some cases there is not enough long wavelength information to fit a modified exponent in which case $\beta$ = 1 and $\lambda_{0}$ = 210~$\mu$m are adopted by default.

\begin{equation}
B_{\rm modified} = B(\lambda,T) \times \left( \frac{\lambda}{\lambda_{0}} \right)^{-\beta} \; ; \; when \; \lambda > \lambda_{0}
\label{mbb}
\end{equation}

\begin{table}
\centering
\caption{Table of stellar parameters from SED fits. The stellar luminosity and mass are in solar units.  The effective temperature is in Kelvin. Stellar masses were tabulated from main-sequence stellar models from \protect\cite{SK82} and broadly belived within $\pm0.4$ solar masses given variation seen when comparing these measurements to more modern stellar models \protect\citep{PM13}. \label{tab:stars}}
\begin{tabular}{cccc}
\hline
Star (HD) & $L_{*}$ ($L_{\astrosun}$) & $T_{eff}$ (K) & $M_{*}$ ($M_{\astrosun}$) \\
\hline
~~11413 & 20.5 $\pm$ 0.34 & 7818 $\pm$ 38 & 1.8~ \\
~~30422 & 8.72 $\pm$ 0.17 & 7948 $\pm$ 54 & 1.9~ \\
~~31295 & 14.7 $\pm$ 0.47 & 8666 $\pm$ 95 & 2.3~ \\
~~74873 & 10.8 $\pm$ 0.21 & 8340 $\pm$ 48 & 2.1~ \\
110411 & 13.2 $\pm$ 0.25 & 8835 $\pm$ 58 & 2.4~ \\
125162 & 17.1 $\pm$ 0.31 & 8606 $\pm$ 52 & 2.3~ \\
183324 & 15.7 $\pm$ 0.28 & 8605 $\pm$ 53 & 2.3~ \\
198160 & $23^{\dagger}$ & 7905 $\pm$ 98 & $2.4^{\dagger}$ \\
221756 & 32.2 $\pm$ 0.60 & 8391 $\pm$ 46 & 2.1~ \\
\hline
\end{tabular}\\
\raggedright
\begin{footnotesize}
$^\dagger$ HD198160 has an indeterminate luminosity and mass because of the combined luminosity due to its binarity.
\end{footnotesize}
\end{table}

\begin{table*}
\caption{Table of blackbody SED fitting parameters. Inferred dust temperatures, $\rm T_{bb}$, are from the modified blackbody fit (Eq. \ref{bb1}-\ref{mbb}). The fractional luminosity of the excess emission is given as \textit{f}. In the cases where two fits were required, the ``cold'' component typically has the higher fractional luminosity (and therefore mass) in the system. The stellocentric dust radius is in AU and is calculated from Equation \ref{equ:diskrad}.  These radii measurements are approximations based on blackbody grains and scaled based on the temperature of the excess. $\lambda_{0}$ = 210~$\mu$m and $\beta$ = 1.00 were adopted by default when those values were unconstrained by the data (Eq. \ref{mbb}). \label{tab:disks}}
\begin{tabular}{ccccccccc}
\hline
Star & \multicolumn{3}{c}{``Warm" Component} & \multicolumn{3}{c}{``Cold" Component} & & \\
(HD) & $\rm T_{bb}$ (K) & \textit{f} ($\times10^{-5}$) & $\rm R_{bb}$ (AU) & $\rm T_{bb}$ (K) & \textit{f} ($\times10^{-5}$) & $\rm R_{bb}$ (AU) & $\lambda_{0}$ & $\beta$ \\
\hline
~~11413 & $\cdots$ & $\cdots$ & $\cdots$ & 55 $\pm$ 2~~ & 2.42 $\pm$ 0.33 & 118 $\pm$ 10 & 210 & 1.00\\
~~30422 & $\cdots$ & $\cdots$ & $\cdots$ & 75 $\pm$ 1~~ & 4.51 $\pm$ 0.47 & 41 $\pm$ 2 & ~~71 & 0.85 \\
~~31295 & 182 $\pm$ 42 & 1.55 $\pm$ 1.61 & 9 $\pm$ 4 & 63 $\pm$ 3~~ & 6.09 $\pm$ 0.70 & 74 $\pm$ 6 & 123 & 1.00 \\
~~74873 & 246 $\pm$ 91 &  2.80 $\pm$ 5.01 & 4 $\pm$ 3 & 108 $\pm$ 21~~ & 2.04 $\pm$ 0.40  & 22 $\pm$ 8 & 210 & 1.00\\
110411 & 203 $\pm$ 70 & 1.61 $\pm$ 0.28 & 7 $\pm$ 5 & 68 $\pm$ 13~ & 4.77 $\pm$ 0.56 & ~~60 $\pm$ 22 & ~~41 & 0.81 \\
125162 & 106 $\pm$ 6~~ & 2.95 $\pm$ 1.05 & 28 $\pm$ 3~~ & 37 $\pm$ 5~~ & 1.42 $\pm$ 1.21 & 235 $\pm$ 67 & ~~61 & 1.48 \\
188324 & $\cdots$ & $\cdots$ & $\cdots$ & 87 $\pm$ 2~~ & 1.79 $\pm$ 0.13 & 40 $\pm$ 2 & 210 & 1.00 \\
198160 & $\cdots$ & $\cdots$ & $\cdots$ & 79 $\pm$ 6~~ & 1.98 $\pm$ 0.63 & 41 $\pm$ 6 & ~~71 & 0.49 \\
221756 & $\cdots$ & $\cdots$ & $\cdots$ & 88 $\pm$ 4~~ & 1.50 $\pm$ 0.16 & 57 $\pm$ 5 & 149 & 1.00 \\
\hline
\end{tabular}
\end{table*}

\begin{table*}
\caption{Table of observed stellar velocities. Right ascension and declination are in the J2000 epoch. $\mu_{ra}$ and $\mu_{dec}$ are the proper motions of RA and Dec in milliarcseconds per year.  Parallax is measured in arcseconds. Parallax is converted to distance in parsecs (pc) for reference. All measurements were compiled utilizing SIMBAD for \textit{Hipparcos} and spectroscopic radial velocity data \citep{FL07,GG06}.\label{vel}}
\begin{tabular}{cccccc}
\hline
Star (HD) & $\mu_{\rm ra}$(mas/yr) & $\mu_{\rm dec}$(mas/yr) & $\rm v_{ rad}$(km/s) & Parallax(\arcsec) & Distance (pc) \\
\hline
~~11413 & -48.27 $\pm$ 0.24 & ~~~-4.42 $\pm$ 0.30 & ~~3.0 $\pm$ 0.7 & 12.96 $\pm$ 0.30 & 77 \\
~~30422 & ~~-3.82 $\pm$ 0.23 & ~~~17.58 $\pm$ 0.33 & ~14.4 $\pm$ 1.0 & 17.80 $\pm$ 0.33 & 56 \\
~~31295 & ~41.49 $\pm$ 0.26 & -128.73 $\pm$ 0.16 & ~11.1 $\pm$ 1.2 & 28.04 $\pm$ 0.25 &  36 \\
~~74873 & -64.46 $\pm$ 0.51 & ~~-51.69 $\pm$ 0.29 & ~23.3 $\pm$ 2.0 & 18.53 $\pm$ 0.43 & 54 \\
110411 & ~82.67 $\pm$ 0.20 & ~~-89.08 $\pm$ 0.13 & ~~~1.6 $\pm$ 2.0 & 27.57 $\pm$ 0.21 &  36 \\
125162 & -187.33 $\pm$ 0.14~~ & ~~159.05 $\pm$ 0.11 & ~~-7.9 $\pm$ 1.6 & 32.94 $\pm$ 0.16 & 30 \\
183324 & ~~-1.01 $\pm$ 0.35 & ~~-32.83 $\pm$ 0.22 & ~12.0 $\pm$ 4.3 & 16.34 $\pm$ 0.36 & 61 \\
198160 & ~83.74 $\pm$ 0.45 & ~~-46.35 $\pm$ 0.59 & -16.0 $\pm$ 7.4 & 13.10 $\pm$ 0.64 & 76 \\
221756 & -17.14 $\pm$ 0.17 & ~~-46.69 $\pm$ 0.15 & ~13.1 $\pm$ 0.6 & 12.45 $\pm$ 0.26 & 80 \\
\hline
\end{tabular}
\end{table*}

In the case of HD 31295, HD 74873, HD 110411 and HD 125162, two blackbody functions were fit to the SED in order to account for the mid-IR as well as the far-IR excess.  Each is refereed to as a ``warm" or ``cold" component in Table \ref{tab:disks}. Based on these SED fits, the basic parameters of the dust can be derived. The radius of the dust will scale with a dust temperature and stellar luminosity relation, given that the dust has reached an equilibrium. Using Equation \ref{equ:diskrad}, where $\rm T_{bb}$ is the excess blackbody temperature in Kelvin and $\rm L_{*}$ is in solar luminosity, gives $\rm R_{bb}$ in AU \citep{MW08}.\\

\begin{equation}
\rm{R_{bb}} = \left( \frac{278.3}{\rm T_{bb}}\right)^{2} \sqrt{\rm{L_{*}}}
\label{equ:diskrad}
\end{equation}

\noindent Dust parameters calculated from the blackbody SEDs are listed in Table \ref{tab:disks}. The uncertainties in the dust radius result from propagating the uncertainty in the temperature and luminosity determined in the SED fits. \\

Comparing the derived stellar parameters to other literature values show they are largely comparable within one sigma uncertainties.  A few outliers exist by a few hundred Kelvin in effective temperature and $\sim$0.1 log solar luminosity, but given we more completely sample the Planck function and fit to a stellar model, we feel this is an improvement on past methods of using purely optical photometry to classify the star \citep{EP02}. 

\section{ISM Bow Wave or Debris Disk?}

Given the image analysis in Section \ref{extentemission} and SED fitting in Section \ref{sedsec}, we can now compare the observations of the excess emission with expectations of an ISM interaction model and a debris disk model.  First, we consider the debris disk interpretation of the observations. Second, we test the ability to resolve the bow structure.  Lastly, we test whether the excess emission temperature and radial extent are consistent with ISM or debris disk dust grains given their respective size and composition.

\begin{table*}
\caption{Table of Galactic stellar velocities and model bow wave characteristics. The Galactic velocities (U, V, W) for the target stars are measured by proper motions and line-of-sight velocities from Table \ref{vel}.  All velocities are in km/s. The heliocentric speed through the Galaxy is given by $v_{\rm gal}$. The average relative velocity with a cloud is given as $v_{\rm rel}$, assuming local ISM clouds travel at $\sim$7 km/s relative to the local sidereal rate \citep{AC97}.  It can be seen that in most cases the measurement uncertainty of the stellar motion is much less than the systematic error in estimating the ISM cloud's velocity of order $\pm$7 km/s. The avoidance radius ($r_{\rm av}$) and temperature ($T_{\rm av}$) for astrosilicate composition and 0.1~$\mu$m grains for the ISM are given by Equations \ref{bow1}$-$\ref{tempbb}. \label{gvel}}
\begin{tabular}{cccccccc}
\hline
Star (HD) & U & V & W & $\rm v_{gal}$ & $\rm v_{rel}$ & $\rm r_{av}$ (AU) & $\rm T_{av}$ (K) \\
\hline
~~11413 & -23.01 $\pm$ 0.03~~ & ~22.05 $\pm$ 0.27~ & ~0.35 $\pm$ 0.22 & 31.78 $\pm$ 0.30~~ & 32.6 & ~9 & 261\\
~~30422 & ~2.61 $\pm$ 0.27 & ~8.05 $\pm$ 0.06 & -2.38 $\pm$ 0.68 & 8.79 $\pm$ 0.74 & 11.2 & 25 & 152\\
~~31295 & -3.97 $\pm$ 1.23 & -9.62 $\pm$ 0.07 & -3.33 $\pm$ 0.14 & 10.93 $\pm$ 1.24~~ & 13.0 & 53 & 137\\
~~74873 & ~13.60 $\pm$ 1.93~ & -8.31 $\pm$ 1.54 & ~0.95 $\pm$ 0.55 & 15.97 $\pm$ 2.53~~ & 17.4 & 12 & 248\\
110411 & -28.64 $\pm$ 0.07~~ & ~~~8.09 $\pm$ 2.87~~ & ~4.09 $\pm$ 1.05 & 30.04 $\pm$ 3.06~ & 30.8 & 7 & 305 \\
125162 & ~26.85 $\pm$ 0.02~~ & ~7.10 $\pm$ 1.22 & 1.60 $\pm$ 1.33 & 27.82 $\pm$ 1.80~ & 28.7 & 5 & 350 \\
183324 & -22.72 $\pm$ 10.98 & ~13.87 $\pm$ 2.70~~ & ~0.80 $\pm$ 4.82 & 26.63 $\pm$ 12.30 & 27.5 & 12 & 234\\
198160 & ~27.26 $\pm$ 27.92 & ~~~5.21 $\pm$ 21.97 & -3.98 $\pm$ 4.94 & 28.04 $\pm$ 35.87 & 28.9 & 23 & 200\\
221756 & -24.93 $\pm$ 0.07~~ & ~0.25 $\pm$ 0.26 & -2.89 $\pm$ 0.03 & 25.09 $\pm$ 0.28~~ &  26.1 & 10 & 227\\
\hline
\end{tabular}
\end{table*}

\begin{figure*}
\centering
\begin{tabular}{ll}
\subfigure{\includegraphics[width=0.48\textwidth,clip,trim=8mm -1cm 20mm 0mm]{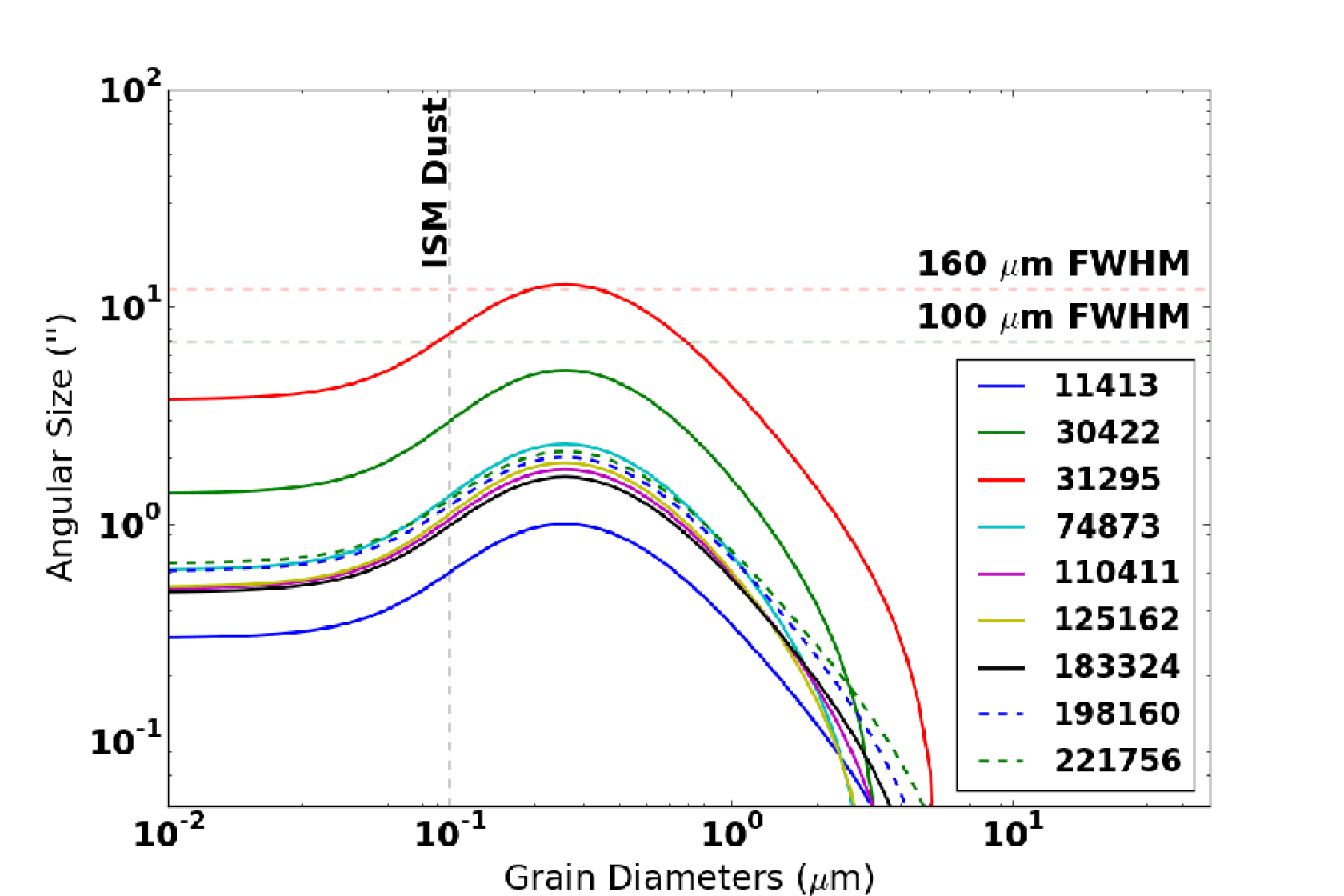}}\put(-70,150){\textit{\textbf{Silicate-Organics}}} &
\subfigure{\includegraphics[width=0.48\textwidth,clip,trim=8mm -1cm 20mm 0mm]{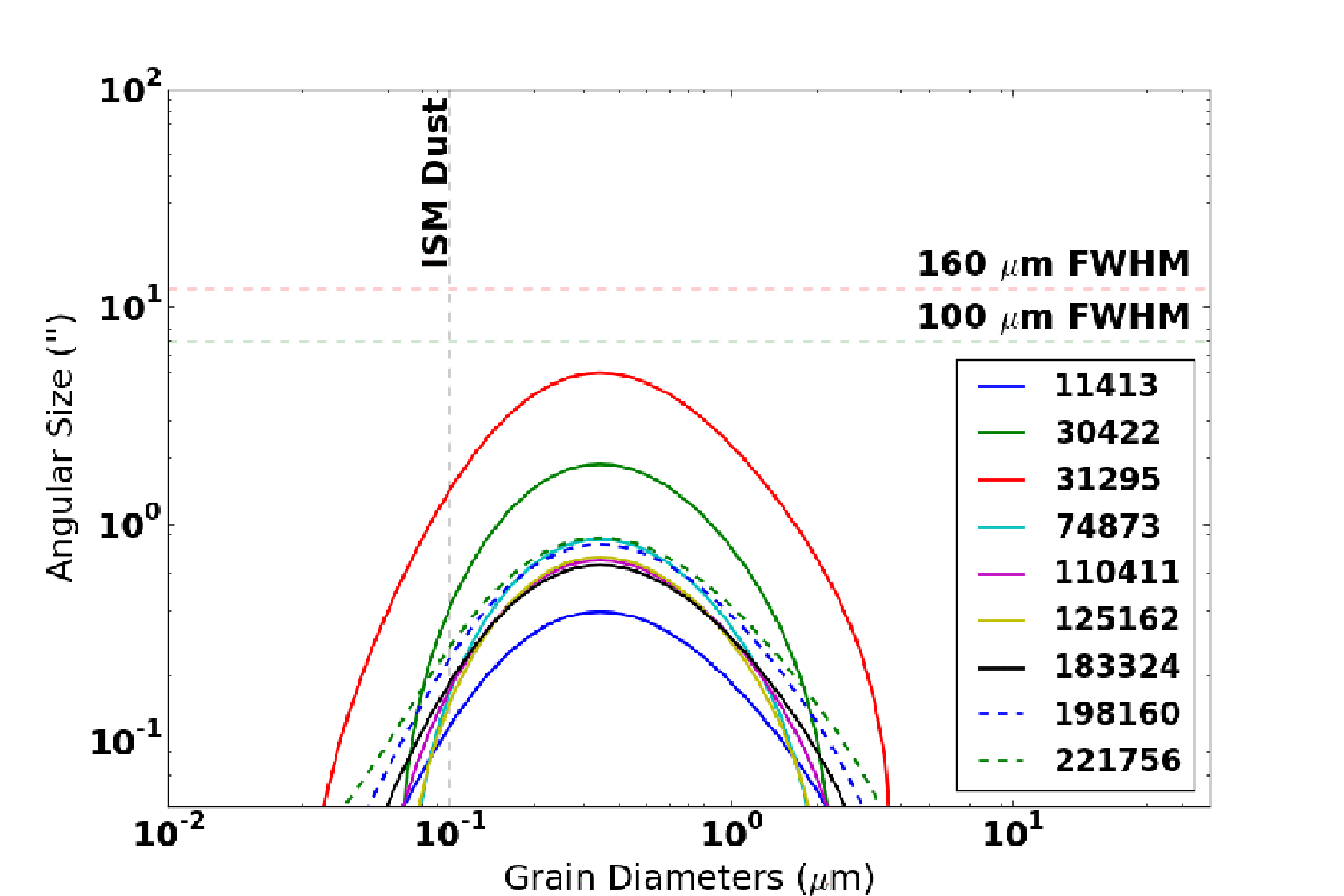}}\put(-53,150){\textit{\textbf{Astrosilicate}}}  \\
\end{tabular}
\caption[Bow Shock Apex Size]{The angular size of the avoidance radius as a function of dust grain size given a silicate-organic composition (Left) and astrosilicate composition (Right).  Typical ISM dust grain size of 0.1~$\mu$m is marked with a vertical line.  The FWHM of the \textit{Herschel} PACS beam at 100 and 160~$\mu$m are shown for reference as horizontal lines. HD 31295 has a bow wave model with the highest potential to be resolved. \label{fig:bow_models}}
\end{figure*}

\subsection{Debris Disk Interpretation}

Many observations such as these have been found to be the result of debris disks, however in this study we must not assume this to be the case given we are trying to differentiate the potential mechanisms of the Lambda Boo phenomenon, even if circumstellar disks are predominately the default interpretation. In most cases, debris disks are parameterized by a single temperature due to the dust typically being arranged in a narrow annulus around the star.  For a two-body fit, it is commonly presumed and sometimes confirmed, that a second, inner belt exists within the system \citep{GK14}.  This would not be unrealistic given our solar system is a two-belt system with a warm asteroid belt between Mars and Jupiter's orbits and a cold Kuiper belt beyond Neptune's orbit.  Since these SED fits of the inner disks are not confirmed or resolved disks (whereas the outer disk are in some cases), we can only say that the blackbody SED fits are consistent with two separate components.  To a first order approximation, we can consider the blackbody temperatures and radii to be the debris disks temperatures and radii.  The blackbody radius is typically underestimated due to the specific dust grain properties, which only serves to make them more resolvable with Herschel.  Since the on sky separation is greater than the instrumental FWHM for a few of the stars we can say their outer radius of emission is resolved. 

As an example HD 31295 has a resolved outer radius which is 129 AU with a blackbody SED estimate of 74 AU.  This is a factor of 1.74 increase in distance from the blackbody radius and is consistent with previously resolved debris disks ranging from 1 to 2.5 times the blackbody radius, thus making its dust properties consistent with other debris disk systems \citep{MB13}.  The precise size distribution of dust and compostion will lead to this offset from blackbody grains. All of the stars have ``cold" component $R_{bb}$ within a factor of 2.5 of the $R_{outer}$ as measured at 100~$\mu$m, except for HD 74873.  Still HD 74873 is within a factor of 4, which can result from late spectral type stars or variations in size distributions of the dust \citep{MB13}.  Previously HD 125162 was fit with a single exponent but was noticeably left with excess flux in its residual after an annulus disk model was applied, suggesting there is some warm component or the structure of the disk is more widely distributed than a simple ring \citep{MB13}.

HD 198160 is the most consistent with an unresolved point source.  This may be due to the binary companion truncating the outer edge of a circumstellar disk around one of the stars. The binary pair is separated by 2.4 arcseconds (or 182 AU projected on the sky) and is a resolved pair of equal magnitude stars of likely the same mass \citep{DJ99}.  If the emission were the result of a bow wave, the ISM cloud could encompass both stars (e.g. the multiple component system $\delta$ Velorum; \citep{AR02}) since it would likely be larger then 182 AU. A circumbinary disk would be unlikely as the configuration would be unstable from an oscillating gravitational potential. The stellar separation and inferred dust radius from temperature match observed properties of a stable circumstellar disks in binary star systems \citep{DR15}. This provides ancillary, but circumstantial evidence that the excess emission observed here is from a debris disk and not a bow wave. The total ISM cloud would need to be more compact than 182 AU and not be influenced as the stars orbit their mutual centre of mass.  The unresolved nature of the PACS data means we cannot explicitly determine the true structure of the emission, as the potential bow waves structure can be smaller in scale than the resolution limit.

\subsection{Bow Wave Models}
\label{bowmodels}

When a star passes through a pocket of ISM dust, it creates a bow wave like structure in the direction of its relative motion.  Radiation pressure creates a cavity of avoidance within the cloud where the dust is repelled by the radiative force from the star.  \cite{AC97} developed the following model for an ISM bow wave based on the required physics described by Equations \ref{bow1} and \ref{bow2}.

\begin{equation}
\rm {r_{av}}(a) = \frac{2(\beta(a)-1) \; M_{*} \; G}{v^{2}_{\rm rel}}
\label{bow1}
\end{equation}

\begin{equation}
\beta(a) = 0.57 \; Q_{\rm pr}(a) \; \frac{\rm L_{*}}{\rm M_{*}} \; a^{-1} \; \rho^{-1}
\label{bow2}
\end{equation}

The avoidance radius ($r_{\rm av}$(a)) is the bow wave's apex or the closest a dust grain of a given size can get to a star with an impact parameter of zero.  While this is the location of the peak brightness, due to its proximity to the star, the cloud as a whole will be irradiated and should have extended surface brightness farther from the star as well. This avoidance radius is a function of the ratio of solar radiation pressure pushing the dust outward relative to the gravitational force pulling it in \footnote{Not to be confused with exponent $\beta$ of Equation \ref{mbb} commonly used in SED fitting}($\beta$(a)) and the velocity of the star through the cloud ($\rm v_{rel}$).  The avoidance radius is proportional to the inverse square of the relative velocity, such that a faster moving star will compress a wave front closer to the star itself.  The parameters of the star such as mass ($\rm M_{*}$) and luminosity ($\rm L_{*}$) relative to solar can be determined by SED fits shown previously (Table \ref{tab:stars}). Assumptions about the dust grains such as size ($a$) in $\mu$m, density ($\rho$) in g cm$^{-3}$, and absorption efficiency ($Q_{\rm pr}$) determine how effective the radiation pressure is, using Mie theory. 

Since all of our target stars are bright and nearby, \textit{Hipparcos} measurements of proper motions have been well determined \citep{FL07}.  Radial velocities along the line of sight have also been measured from offsets in spectroscopic line measurements \citep{GG06}.  These measurements are compiled in Table \ref{vel}.   Using these velocities, the actual motion of the star within the Galaxy can be calculated using a matrix transformation by knowing the location of the Galactic centre and the projection of velocities relative to earth \citep{JS87}. U, V, and W are all positive towards Galactic anti-centre, mean Galactic rotation, and North Galactic pole, respectively. The effect of the observer's motion is removed by subtracting the local sidereal velocity \citep{CB11}.  The final galactic relative velocities, $\rm v_{gal}$, are calculated in Table \ref{gvel}. Other literature sources calculated galactic velocities for these stars, but did not use contemporary \textit{Hipparcos} measurements and didn't correct for the local sidereal rate necessary to compare with ISM measurements \citep{EP02}.

\begin{figure*}
\centering
\begin{tabular}{ll}
\subfigure{\includegraphics[width=0.48\textwidth,clip,trim=8mm -1cm 20mm 0mm]{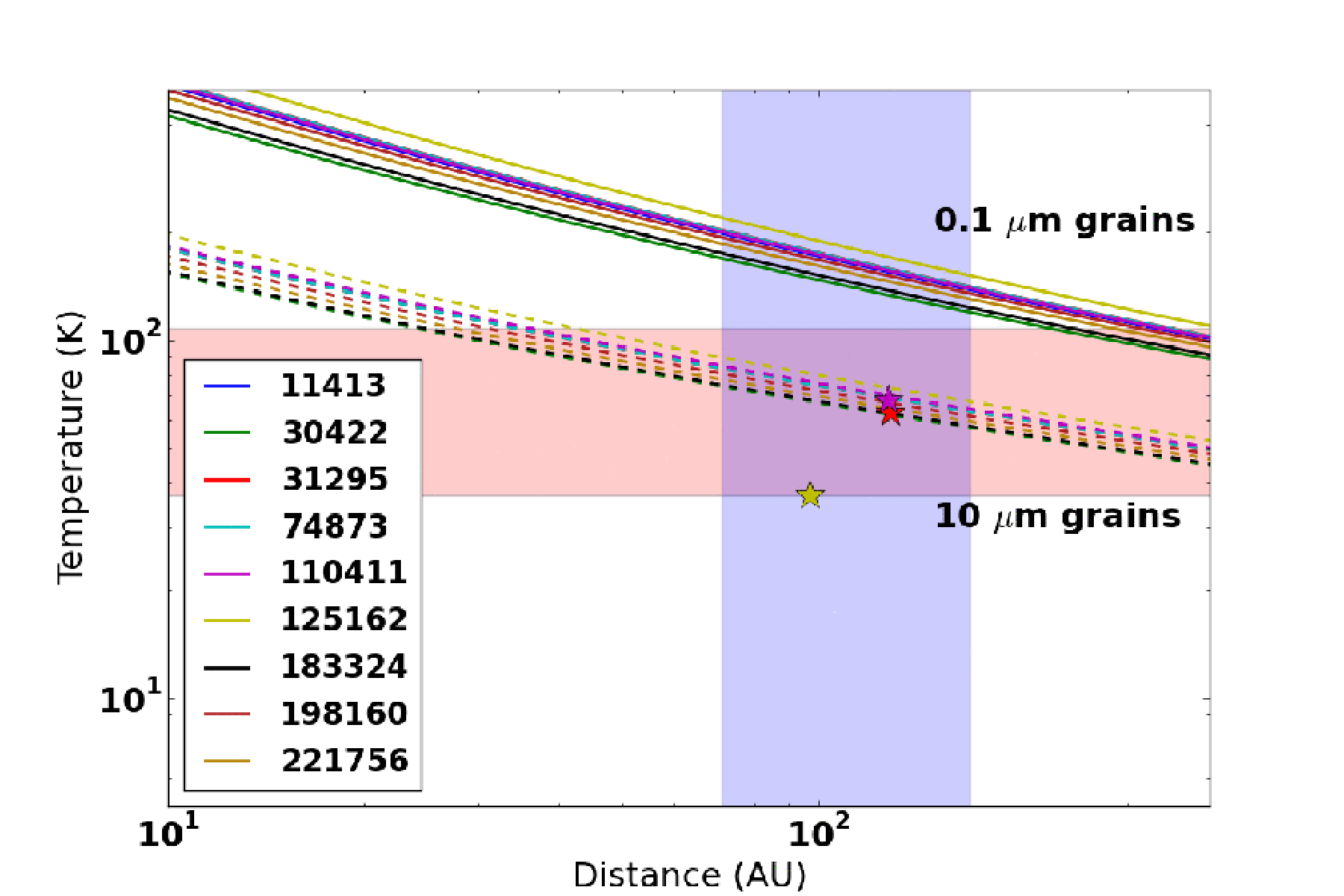}}\put(-70,150){\textit{\textbf{Silicate-Organics}}} &
\subfigure{\includegraphics[width=0.48\textwidth,clip,trim=8mm -1cm 20mm 0mm]{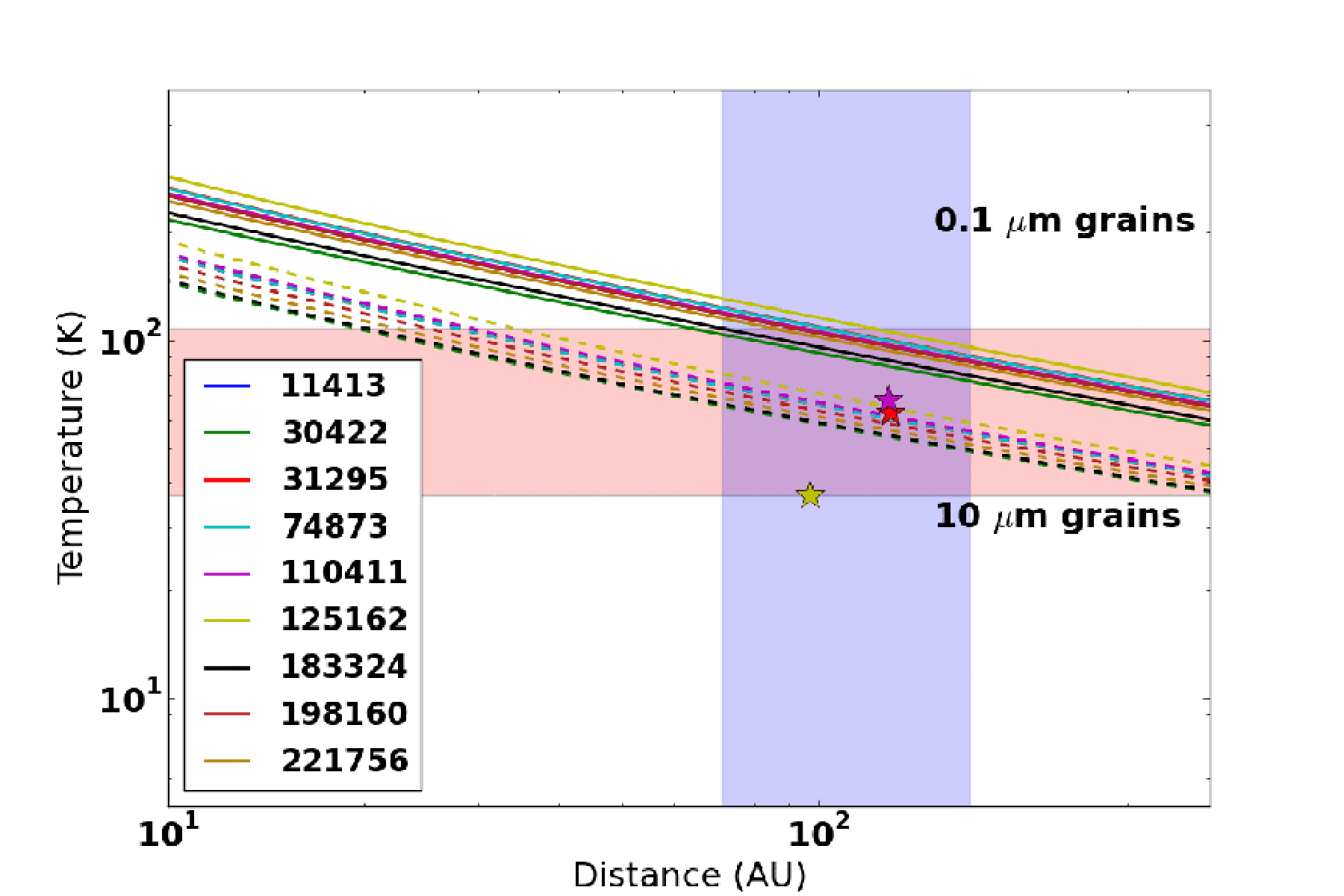}}\put(-53,150){\textit{\textbf{Astrosilicate}}} \\
\end{tabular}
\caption[Temperature vs Distance]{Plots showing Equation \ref{dusttemp} for 10~$\mu$m grains (dotted), typical of debris disks around A-type stars, and 0.1~$\mu$m grains (solid) typical of the ISM. On left, a dust composition of silicate-organics is used in calculating the absorption efficiency for dust. On right, an astrosilicate composition is used. The temperature range based on the ``cold" component SED fits from Table \ref{tab:disks} (37-108 K) is shown in red.  The outer radial extent ($R_{outer}$) of the 70/100~$\mu$m emission from Table \ref{angles_100} (71-171 AU) is shown in blue.  Stars denote the exact measurments of the resolved stars HD 31295, HD 110411, and HD 125162.  The overlapping region in purple is the parameter space consistent with observations.  Error bars have been included on the total range, except systems which are at the PACS resolution limit.  Debris disk size dust matches the temperature and radius well regardless of composition.  Astrosilicate composition of ISM size dust is more well suited to the observations but still falls outside the measured temperature range for the given distance.  Bow wave models would suggest that the dust is located much closer, at 5-53 AU, which would mean the dust grains would be hotter than what is observed in red. \label{eqtemp}}
\end{figure*}

Since the ISM cloud itself can also have velocity relative to the star, we add in quadrature an additional velocity term of 7 km/s as an estimate of the actual cloud to star relative velocity or $\rm v_{rel}$ \citep{AC97}.  The cloud could be as much as $\pm$7 km/s but would require precise alignment of the two velocity vectors which is unlikely. In most cases the stellar velocity is much higher than the ISM velocity and therefore only modestly affects the cloud-star relative velocity (see Table \ref{gvel}).  

Furthermore, we assume $\rho$ is 3.3 g cm$^{-3}$ as the dust density typical of the ISM \citep{DR84} and used in previously observed and modelled bow waves \citep{AC97,GA08}. In conjunction with the mass and luminosity from the SED fits, $\beta(a)$ can be calculated using Mie theory and an assumed composition as a function of grain size. Since $Q_{\rm pr}$ can vary with composition, we calculated $\beta(a)$ for two compositions; pure astronomical silicate \citep{DR84} and a mix of silicates and organics \citep{AU99}.  Astrosilicates are typical of the ISM, while silicate organics are typical of debris disks \citep{AU99}.  This in turn gives the avoidance radius of the bow wave from the star via Equation \ref{bow1}.  This radius is divided by the distance (d) to the star to get the angular size and is compared with the PSF FWHM of \textit{Herschel} PACS as a function of grain size in Figure \ref{fig:bow_models}. 

It can be seen in Figure \ref{fig:bow_models} that the system which has the best chance for a bow wave to be resolved is HD 31295.  Since HD 31295 is well resolved to have symmetric features not aligned to its motion at both 100 and 160~$\mu$m, we have some evidence that HD 31295 is not interacting with an ISM dust cloud (see Figure \ref{angles_100}).  For the other stars, \textit{Herschel} cannot explicitly resolve a bow wave apex.

Other studies on ISM bow waves, such as by \cite{JM09}, calculate the outer radius of a uniform density ISM cloud heated by the star needed to be consistent with the excess 70~$\mu$m emission from \textit{Spitzer} observations. They integrate a size distribution from 0.001 to 10 microns using models consistent with the ISM from \cite{DR84}.  These are also the same type of models shown in Figure \ref{PACS_models}. For the 4 stars in common between these studies, the outer regions of excess emission were found to have to be $\sim$1500 AU away from the star. We can constrain the observed emission to within 150 AU for all stars (see Section \ref{extentemission}). This both rules out an ISM bow wave model and rules out confusion with nearby cirrus ISM emission, as was observed with false positive debris disks from IRAS \citep{PK02} or what is seen in the background within the FOV of HR 8799 with \textit{Herschel} \citep{BM14B}. Therefore we again favour a debris disk model

This does not rule out debris disk models where emission is dominated by larger 10~$\mu$m grains due to radiation pressure propelling out smaller grains. In fact, such large grains will not be present in a bow wave as the radiation pressure is too inefficient to divert the dust grains (i.e. $\beta$(10~$\mu$m) $<$ 1), which can be seen in the rapid drop of $\rm r_{av}$ in Figure \ref{fig:bow_models}.

\subsection{ISM vs Debris Disk Equilibrium Temperature}
\label{equtemp}
			
By combining the derived temperatures and spatial scale, it is in principle possible to distinguish between the two models of an ISM bow wave and a debris disk using grain size. Typical dust grains in the ISM are of order $\sim$0.1~$\mu$m due to its origin in AGB winds and supernova \citep{DR84}. Evidence of this is seen in the size distribution of ISM grains peaking at 0.1~$\mu$m grains \citep{JM77,JM96,SK94}. Also, meteorite samples have shown pre-solar grains are typically less than a micron, but larger grains of a few microns can still be found \citep{AD11}. On the other hand, disk grains are typically $\sim$10~$\mu$m in size for A-type star luminosities as the blow-out grain size is on the order of a few microns \citep{AU99}. This means grains smaller than $\sim$10 microns will be ejected from the system by radiative pressure on hyperbolic orbits, leaving behind the larger grains \citep{BLS79}. 

The variation in grain size leads to a change in the equilibrium temperature of dust at a given stellocentric radius from the star. Dust grains are less efficient emitters at wavelengths much greater than the grain size \citep{GB94}. The smaller grains will therefore reach a higher equilibrium temperature than larger grains.  The temperature of a dust gain of size, $\rm s$, at a radius from the star, $\rm r$, is given below \citep{GB94}:

\begin{equation}
T(a,r) = \left( \frac{<\!\!Q_{abs}\!\!>_{T_{*}}}{<\!\!Q_{abs}\!\!>_{T(a,r)}} \right)^{0.25} T_{\rm bb}
\label{dusttemp}
\end{equation}

\begin{equation}
T_{\rm bb} = \frac{278.3}{\sqrt{r}} L_{*}^{0.25}
\label{tempbb}
\end{equation}

\noindent$<\!\!Q_{abs}\!\!>$ is the absorption efficiency averaged over either the stellar spectrum (denoted with $T_{*}$) or the blackbody spectrum at a given dust temperature (denoted with $T(a,r)$). Since $T(a,r)$ is on both sides of the equation it requires iterative solving to converge the temperature on either side. Solving $T(a,r)$ for both sizes of dust grains (a$=$0.1 and 10~$\mu$m) around each star and plotting as a function of stellocentric distance allows comparison with the measurements of the temperatures and outer radial extent of the excess emission. The absorption efficiency is again related to composition so we use the same compositions from Section \ref{bowmodels}. The computed temperature-distance curves for each star can be seen in Figure \ref{eqtemp}.

Since \textit{Herschel} PACS data have higher resolution at 70/100~$\mu$m and has less contamination from background sources, we use those radii measurements to constrain the radial scale of the excess emission. For constraining temperature, we use the range of blackbody temperatures from the ``cold'' components in Table \ref{tab:disks}. In Figure \ref{eqtemp} (Left), a mixture of organic silicates was used as the composition for determining the absorption efficiencies, which is typical of debris disks \citep{AU99}. In Figure \ref{eqtemp} (Right), a mixture of astrosilicates typical of the ISM was used \citep{DR84}. It can be seen that given 0.1~$\mu$m dust grains, the curves are outside the observed temperatures and radii (purple region). Larger grains, however, cross through the region constrained by the measurements of the excess emission regardless of composition. Astrosilicates which are more typical of the ISM are slightly closer to the measured values but are still warmer than the emission observed.   Of the two compositions, silicate organics best fit the resolved excess measurements denoted as stars in the figure. It is also important to note that bow wave models show that 10~$\mu$m grains will not be a major constituent of dust in the bow wave itself. The radiation pressure is too ineffective such that the effective radius of avoidance plummets very close to the star for these grains, indicating they will simply pass-by even if they were present in the ISM (See Figure \ref{fig:bow_models}). We therefore conclude that the excess emission from the stars stems from debris disks, rather than ISM bow waves, because the data are consistent with larger 10~$\mu$m dust which is generally colder for its given stellocentric distance to the host star.  This is further supported by the lack of resolving large scale emission from a cloud.

The temperature of dust at the avoidance radii ($\rm r_{av}$) from the bow wave models in Section \ref{bowmodels} effectively discredit an ISM bow wave of pure astrosilicates because the temperature of 0.1~$\mu$m dust would be $\sim$137-350 K at the minimum modelled radius, which is too high compared to the measured SED temperatures of 50-108 K (see Tables \ref{tab:disks} and \ref{gvel}). While some stars have ``warm" SED components, from $\sim$106-246 K, that are consistent with the bow wave temperatures at $\rm r_{av}$, not all the stars in our sample do. In our sample $44^{+16}_{-14}$\% of disks have two components which is consistent with other estimates for debris disk star hosts in general of order $33\%$ \citep{GK14}. HD 125162 ($\lambda$ Boo) for example, has a warm component SED temperature of 106 K for its disk, but has the highest expected bow wave dust temperature at 350 K. Only through resolved imaging of the inner component can a disk+ISM scenario be completely ruled out. A combination of both are not mutually exclusive since sandblasting (i.e., the erosion of a debris disk from an ISM interaction) will not significantly destroy the debris disk \citep{AC97}.

\section{Correlation with IR-Excess}
\label{corr_excess}

The Lambda Boo phenomenon has often been associated with an IR photospheric excess \citep{EP04}. Estimates of the fraction of Lambda Boo stars which have an IR excess have previously been shown to be $23^{+10}_{-6}\%$ \citep{PE03}, which is typical of A-stars in general \citep{NT14}. The Paunzen et al. estimate is rather conservative and if modified by including disks that may have an excess (a detection in only one band) and excluding stars with only ISO upper limits (which are non-deterministic of an excess), the estimate can be up to $53^{+12}_{-12}\%$ (from 6/26 to 8/15). In fact the system HD 11413, which was previously contentious, is definitively associated with an excess through \textit{Herschel} observations. Therefore the constraints placed in the past are not as conclusive given that more sensitive far-IR observations can more readily detect cold disks around these stars.

While all the Lambda Boo stars presented here exhibit an excess with \textit{Herschel}, this fact is not statistically significant because they were targeted with prior knowledge of their excess. If instead we look at the 123-star \textit{Spitzer} sample (Malmquist unbiased by observing down to the stellar photosphere) from \cite{KS06}, excluding B-type stars, an excess around 40 stars was detected ($\sim$33$\pm3\%$). This sample was only biased by reliable age determinations.  Of the 123 total, there were 13 Lambda Boo stars observed.  Of the 13, 10 have an IR excess detected at 24 or 70~$\mu$m ($77^{+7}_{-14}\%$). Specifically, HD 319, 142703, and 210111 had no excess detections out to 70~$\mu$m, while HD 11413, 30422, 31295, 110411, 111786, 125162, 188324, 198160, 204041, and 221756 had detections in at least one band.  The median age of the sample as a whole is 300 Myr and the median age of the Lambda Boo star sample is also 300 Myr. The median age of Lambda Boo stars with non-detections is higher than it is for detections (600 vs 200 Myr), which may mean non-detections are the result of intrinsically fainter debris disks given debris disk fractional luminosity fades with age \citep{MW08}.  The median distance of the sample as a whole was 74.8 pc while the Lambda boo stars' median distance is 59 pc.  The bias in distance may allow for dimmer debris disks around Lambda Boo stars to be detected more easily than the control sample. For the non-detections around Lambda boo stars, the distances are between 52 and 80 pc.  In general, there are no extreme biases which would clearly account for the discrepant detection rates around Lambda Boo stars (77 vs 33$\%$).  The spectroscopic surveys for identifying Lambda Boo stars do not appear to be biased towards stars with IR-excesses either \citep{GC02}.  

A Fisher Exact test comparing those two populations sets of A-stars and Lambda Boo stars from \textit{Spitzer} results in a p-statistic of 0.0042.  It is therefore very improbable that the two samples have an identical distribution of bright IR excesses.  It is then significant to say that Lambda Boo stars are more likely to have brighter IR-excesses.  Based on our \textit{Herschel} data, we can say that the IR-excesses around Lambda Boo stars most likely arises from a debris disk. Thus, it would logically follow that the higher incidence of bright IR-excess around Lambda Boo stars is really a higher incidence rate of bright debris disks.  Surveying more Lambda Boo stars with a similar or better sensitivity of far-IR observations will strengthen the correlation found here.  Furthermore, we would also need more spectroscopic classification of Lambda Boo stars with such IR observations.

\section{Spectroscopic Composition}

\begin{figure}
\begin{center}
\includegraphics[width=0.49\textwidth]{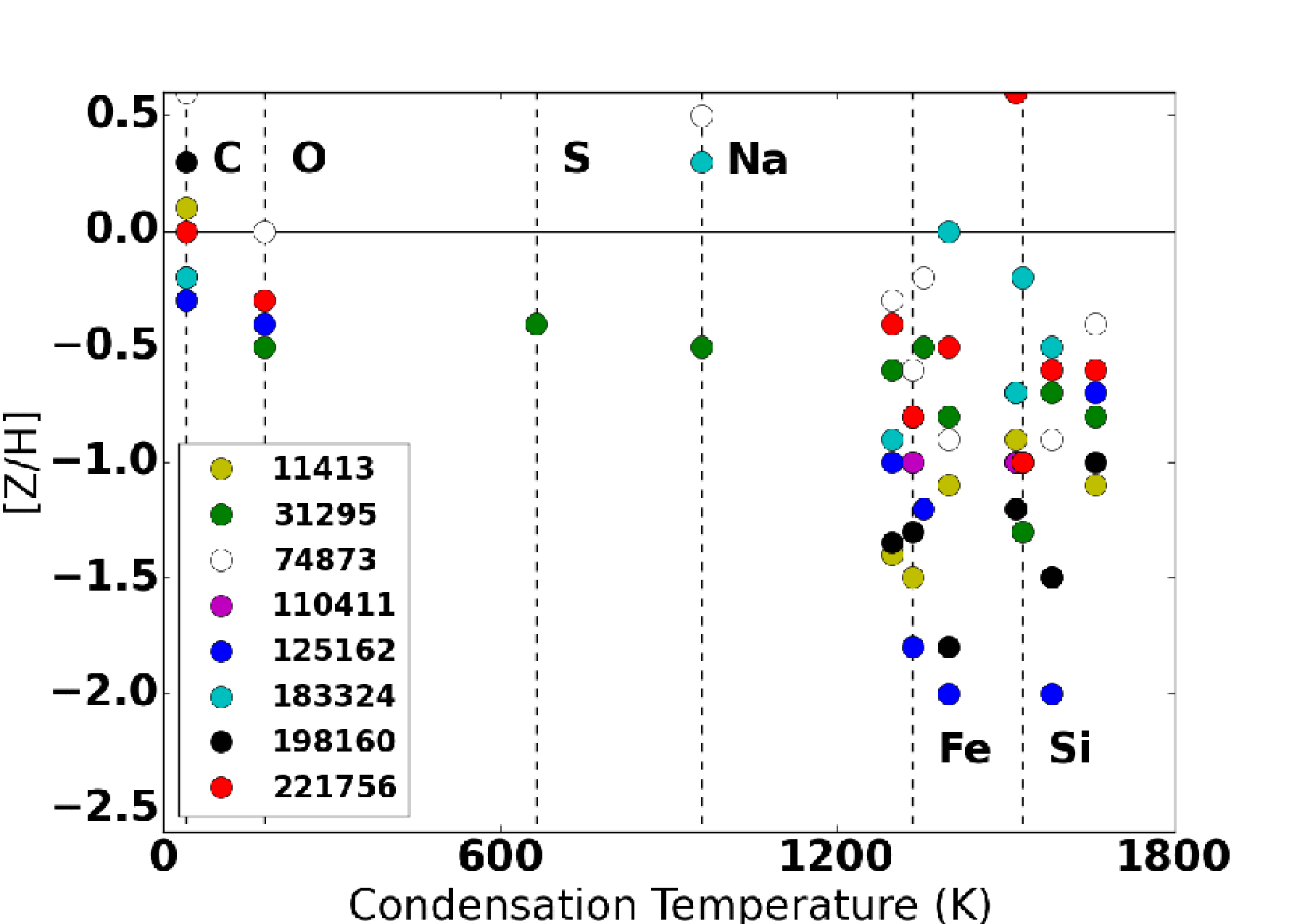}
\caption{\label{condtemps}The metal abundance correlation to sublimation temperature.  The solid horizontal line indicates solar composition. C and O have the lowest condensation temperatures (for elements shown here) and are solar abundant. While $\alpha$ and Fe-peak elements, which are under-abundant, have high condensation temperatures.  Abundances used here typically have measurement uncertainties of $\pm$0.2 dex.  HD 30422 presently has no known refractory abundance measurements and is therefore excluded from this figure, but it has been confirmed as a Lambda Boo star through spectral classification \citep{GC93}.}
\end{center}
\end{figure}

\begin{figure}
\begin{center}
\includegraphics[width=0.49\textwidth]{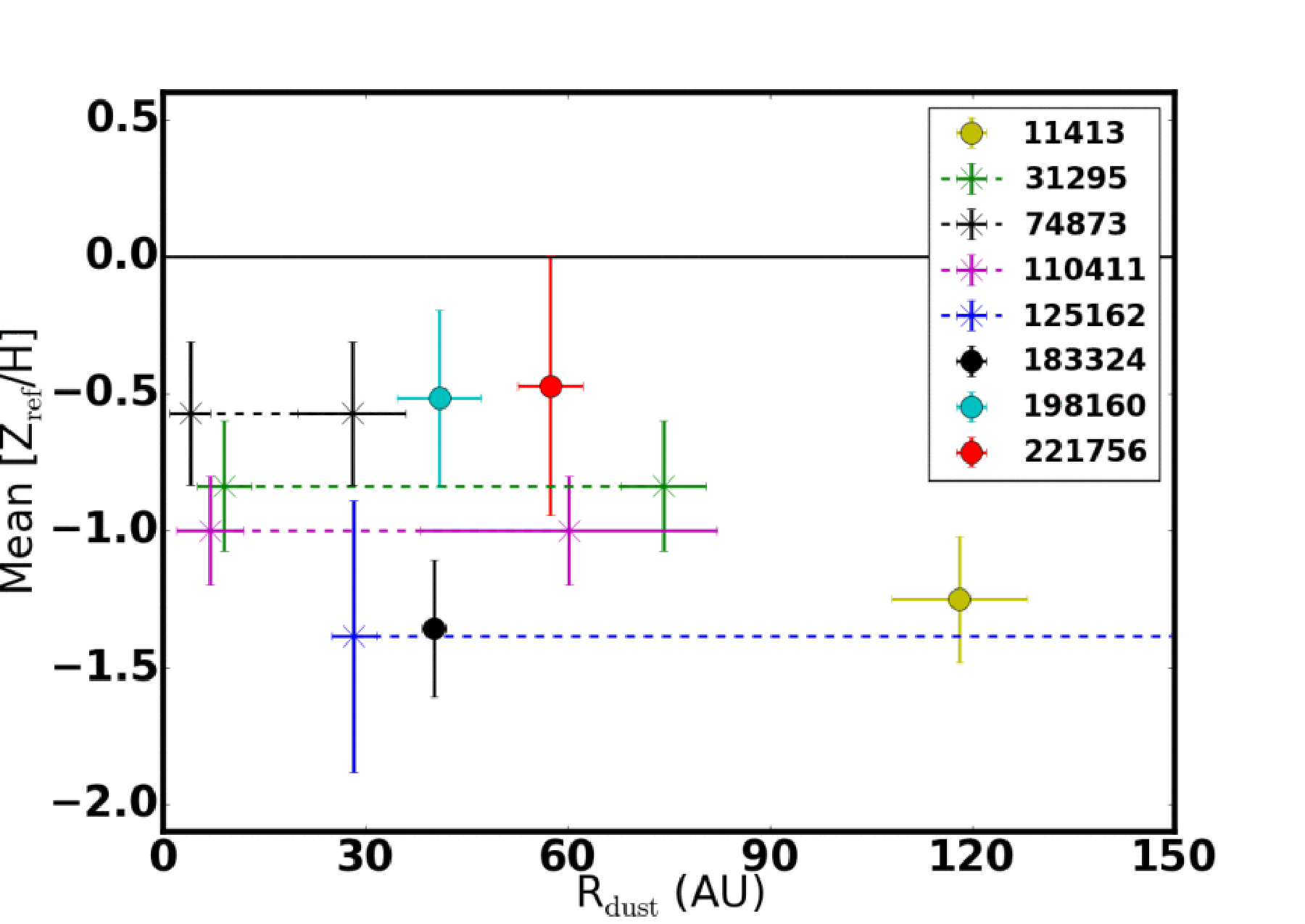}
\caption[]{\label{diskrad}
The mean abundance of refractory elements versus disk radius.  There is a weak trend , if at all, for the lower refractory abundances to be positively correlated with smaller debris disks radii (ignoring HD 11413 as an outlier).  The error in abundance is taken as the standard deviation in the metal abundances.  The error in the disk radius is from the SED fit (see Table \ref{tab:disks}). Crosses show stars with additional ``warm" components.  Dust accretion models such as PR-drag are expected to be radius dependent and therefore potentially correlated (see Section \ref{prdrag}).}
\end{center}
\end{figure}

If there is a causal link between an IR excess (i.e., a debris disk) and the abundance anomaly on the surface, there should be correlations with the dust excess. Logically, there would be a continuum of accretion rates from the different debris disk configurations, which would lead to a variation that is proportional to the relative abundances on the surface.  Figure \ref{condtemps} illustrates a known correlation between volatile and refractory elements using literature values for spectroscopic abundances \citep{HU02,SS93} and elemental condensation temperatures \citep{KL03} for the Lambda Boo stars in this sample. It can be seen that the metal deficiency is: a) present with elements which have condensation/sublimation temperatures greater than 1200 K; and b) the refractory metals which are underabundant vary from star to star.  This relation suggests that the Lambda Boo phenomenon involves dust sublimating in the terrestrial zone of at least a few 100 K, but not too close to the stellar surface such that refractory elements will vaporize or accrete directly on the star. Stars can also have a varying degree of refractory-poor metallicity rather than a strict abundance fingerprint, such that the composition of accreting material must vary from star to star.  The strength of the anomaly may then be strongly related to the proximity of the dust to drive a higher magnitude of accretion. As an example, $\beta$ Pic has a disk where collisions are creating sub-micron dust \citep{CT05} and volatile gas \citep{WD14}, but is not a confirmed Lambda Boo star \cite{HH97}. However, one could imagine that $\beta$ Pic may have been a Lambda Boo star if its disk were closer to the star where the conditions could be met for the volatile gas to viciously accrete onto the star \citep{RF06}. \\

As a test, we try to determine these possible trends with disk configuration. Using the average of the refractory abundances with a condensation temperature greater than 1200 K, the spectroscopic composition of Lambda Boo-like properties and disk properties derived in our sample can be compared. Disk/stellar luminosity, temperature, and mass were not found to have a significant correlation with the mean refractory abundances. Since surface abundances are not correlated with stellar properties, within the context of them all being A-type stars, it is suggestive that it is not an internal mechanism. One parameter which may have some significance is disk radius seen in Figure \ref{diskrad}. If valid, this relation would imply that the closer the dust is observed to the star, the more significant the abundance anomaly. However not all are resolved disk annuli and therefore prone to uncertainty by up to a factor of 2.5 \citep{MB13}. Furthermore, the range in refractory abundances are typically a factor of $\pm$0.4 dex in standard deviation, which doesn't allow for a large degree of variance between Lambda Boo stars given uncertainty in current measurements. It may therefore be possible that stars with bright debris disks do not have the spectroscopic precision to detect a weak incidence of Lambda Boo-like properties. Higher precision spectroscopic measurements and resolved disk imaging are required to observe such trends.

\section{Mechanisms for Secondary Accretion}
\label{prdrag}

\subsection{Poynting-Robertson Drag}

Poynting-Robertson drag is the mechanism by which orbiting dust grains lose momentum and spiral in towards the star.  When dust is being slightly irradiated in the direction of motion, due to its orbital path, the radiation imparts a ``drag" force causing momentum loss from incident photons \citep{BLS79}. \cite{VL14} and \cite{MW05} have worked out several analytic approximations for the accretion rate of dust due to PR-drag into the inner solar system by a collisionally active debris disk.  In general, the model is ideal for explaining the differentiated accretion needed to explain the Lambda Boo phenomenon.  Dust is accreted from a debris disk which acts as a reservoir.  The dust enters the inner stellar system as large grains where it begins to sublimate volatile elements into gas, which is accreted onto the star.  The dust grains will then decrease in size as they sublimate their volatile mass.  The smaller, refractory metal rich dust grains are then more susceptible to radiation pressure and are blown out of the system (see Figure \ref{cart_pr}).  The maximum accretion rate in units of $M_{\Earth}/\rm{yr}$ of dust grains with a given $\beta$ value down to a radius ($\rm r$) of zero is as follows:

\begin{figure}
\begin{center}
\includegraphics[width=0.45\textwidth]{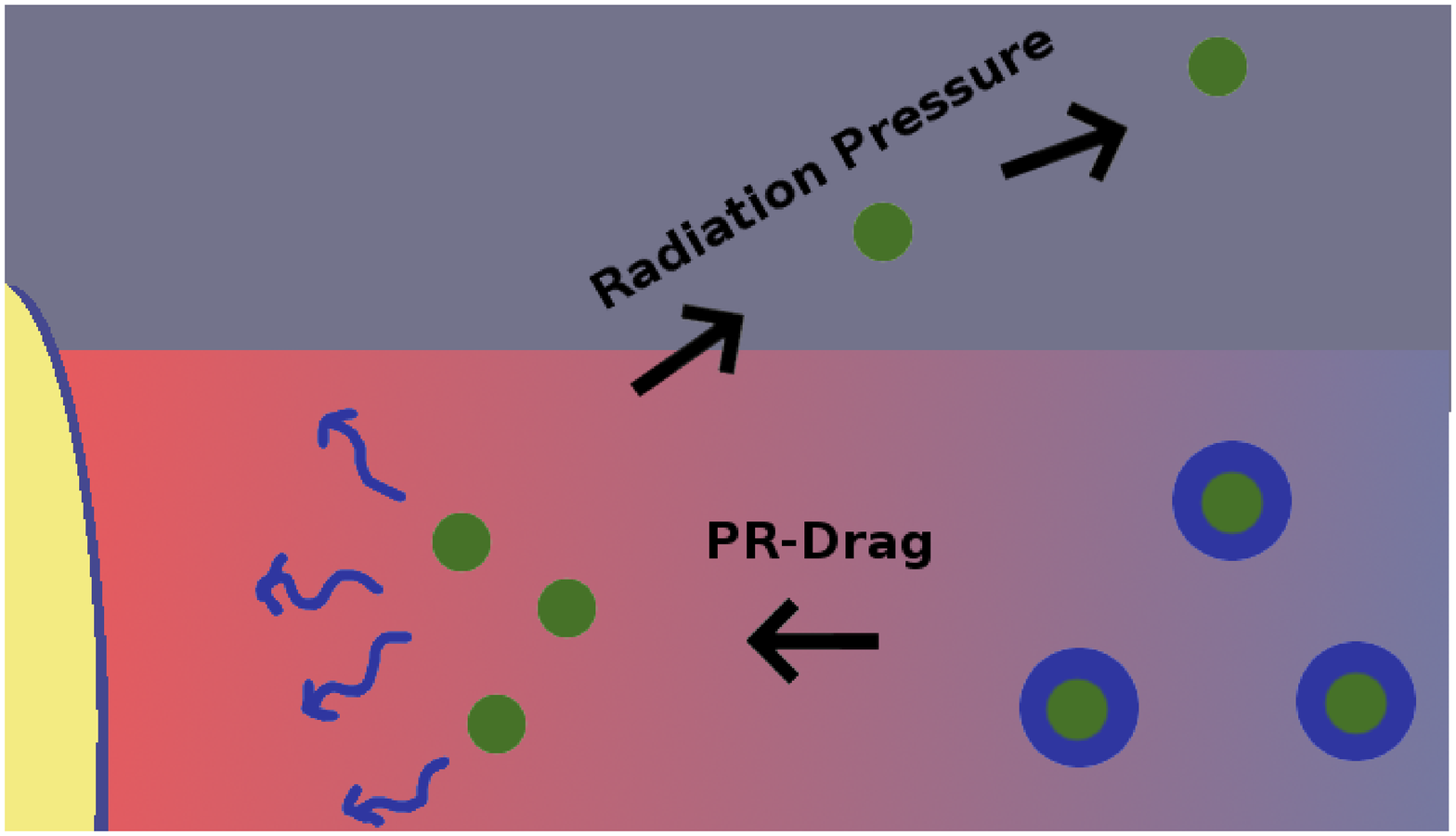}
\caption[]{\label{cart_pr}
A cartoon depicting a model for secondary accretion via a debris disk.  Large grains with volatile elements C, N, O, and S are frozen out on grains (blue).  When PR-Drag brings them into the inner solar system, they sublimate the volatile ices.  The gas accretes onto the star while the now smaller, refractory-rich dust grains (green) experience a higher radiation pressure relative to gravity ($\beta$ $>$ 0.5) and are blown out on hyperbolic orbits.
}
\end{center}
\end{figure}

\begin{equation}
\rm{max}[\dot{M}_{\rm PR}(r=0)] = 5.6 \times 10^{-13} * \frac{\sqrt{M_{*}} \; L_{*}}{\sqrt{R_{\rm disk}}} \; Q_{pr} \; \frac{\beta}{0.5} 
\end{equation}

\noindent $Q_{\rm pr}$ is the radiation pressure efficiency on the dust and $\beta$ is the ratio of radiation pressure to gravity.  These are fundamental properties of the grain which vary with dust grain size and composition.  However, we will assume that we want the maximum possible accretion rate.  We therefore set $\beta$ = 0.5 (as the maximum possible value for bound grains) and $Q_{\rm pr}$ = 2 \citep[as it can be physically confined between 0 to 2;][]{VL14}.  This simplifies the equation to stellar and disk parameters which have been measured for our target stars:

\begin{equation}
\label{accret_equ}
\dot{M}_{\rm max} = 1.12 \times 10^{-12} * \frac{\sqrt{M_{*}} \; L_{*}}{\sqrt{R_{\rm bb}}}
\end{equation}

Where $M_{*}$ and $L_{*}$ are the mass and luminosity in solar units. $R_{\rm bb}$ is the blackbody estiamte of the disk radius in AU.  Using the measurements from Table \ref{tab:disks} and \ref{tab:stars}, the maximum accretion rates of dust are compiled in Table \ref{accret}.  They lie in the range of $2-10\times10^{-12}\;M_{\Earth}/\rm{yr}$.  Gas-to-dust ratios of comet coma, warmed to levels within our solar system, range from 0.1 to 1 \citep{PS92}.  Again for the maximum realistic accretion rate, these accretion estimates need to be reduced by a factor of 2 for the accretion rate of just volatile gas, assuming the sublimation timescale is negligible. These rates are reasonable since they are lower than $0.33\;M_{\Earth}/\rm{yr}$, where gas and dust will entrain and prevent the differentiation of metals in the inner AU of the system \citep{LW92}.  

\begin{table}
\centering
\caption{The maximum accretion rates of dust due to PR-drag in each system based on stellar and disk measurements of Table \ref{tab:stars} and \ref{tab:disks} using Equation \ref{accret_equ}.  HD198160 is a binary star and therefore has radiation effects from both stars which are not adequately approximated in this model.\label{accret}}
\begin{tabular}{cc}
\hline
Star (HD) & $\dot{M}_{\rm max}$ ($\times10^{-12}$ $\frac{M_{\Earth}}{\rm{yr}}$)\\
\hline
~~11413 & 3.0\\
~~30422 & 1.9\\
~~31295 & 7.9\\
~~74873 & 8.7\\
110411 & 8.7\\
125162 & 5.4\\
188324 & 4.0\\
198160 & N/A\\
221756 & 7.1\\
\hline
\end{tabular}
\end{table}

Based upon stellar atmospheric models, the estimated mass of volatile gas required on the surface of a Lambda Boo star is roughly $0.33 \; M_{\Earth}$ \citep{LW92,ST02,ST93}.  A simple inversion of the maximal accretion rate from PR-drag of $10^{-11}\;M_{\Earth}/\rm{yr}$, shows that it will take 33 Gyr for that amount of gas mass to accumulate on the surface.  This exceeds the age of the universe by a significant margin.  Given that the age of a main sequence A star is at most $\sim$2 Gyr, the minimum accretion rate for that amount of mass would need to be $\sim\hspace{-0.9mm}10^{-9} M_{\Earth}/\rm{yr}$ for the phenomenon to occur at some point within the stars' lifetimes and be observable.  Again this is also based on the assumption that there are no dissipation effects on the stellar surface when in fact there are (e.g. meridional circulation).  The minimum plausible estimate is 2 orders of magnitude higher than the estimated maximal accretion rates for these stars given the PR model. It makes sense to rule out PR-drag as the mechanism which causes the Lambda Boo phenomenon, since this mechanism would be universal to debris disks and so would be in contradiction with the fact that not all bright debris disk hosting stars are detected as Lambda Boo stars \citep{HH95,IK02}.  Therefore, some other rare accretion mechanism must play a role in causing the abundance anomaly, if the phenomenon is indeed related to a debris disk mechanism.

\subsection{Dynamical Activity}

It is also possible that the Lambda Boo phenomenon and bright emission from a debris disk may be causally connected due to their independent correlation to a third phenomenon, such as planetary scattering.  It has been proposed that the source of the solar zodiacal cloud is through the continual disruption of comets \citep{ND10}.  Interferometric surveys have found an occurrence rate of $50^{+13}_{-13}\%$ for hot exozodi dust around A stars \citep{DA13,ES14}.  If a moderately sized planet ($<$ Jupiter mass) were to migrate through a cold debris disk it could potentially achieve a sustained accretion rate of $\sim\hspace{-0.9mm}10^{-9} M_{\Earth}/\rm{yr}$ in mass within 3 AU for over a Gyr to replenish the exozodi \citep{AB14B}. This is within reason to achieve $0.33 \; M_{\Earth}$ of volatile gas based on the stellar age constraint from before.  The accretion rate is dependent upon planetary and disk architecture, but higher rates of $\sim\hspace{-0.9mm}10^{-8} M_{\Earth}/\rm{yr}$ have also been achieved in shorter bursts \citep{AB14B}.  For exozodi, sustained accretion is preferred to explain the high prevalence of hot dust. The rarity of the Lambda Boo phenomenon, however, allows for the accretion rate to be higher and occur for shorter periods of time (i.e. late-heavy bombardment events). The diffusion due to meridional circulation is on the order of $\sim\hspace{-0.9mm}10^{-6} M_{\Earth}/\rm{yr}$ \citep{ST02}.  Meridional circulation could then dissipate the abundance anomaly in 1-2 Myr, which means the heightened accretion would need to have occurred recently \citep{ST02}.  The accretion rate in order to overcome diffusion and build a layer of volatile gas equal to $0.33 \; M_{\Earth}$ within 1 Myr would be $\sim\hspace{-0.9mm}10^{-7} M_{\Earth}/\rm{yr}$.  This is higher than previous models have shown, but could be the result of more extreme dynamical scenarios.  It is also plausible that the dust production was produced relatively close in by some large impact scenario of earth mass bodies, such as with the giant impact theory for the Moon's formation \citep{RC08,JP12}.  Planetary impacts may also help explain older Lambda Boo stars ($\sim$1 Gyr) because the circumstellar disk is expected to diminish over time, whereas planetary impacts could provide stochastic bursts of material at later ages.  The rarity of Lambda Boo stars (2$\%$), with the abundance pattern lasting for 2 Myr until the surface can mix, means that only 10 such events are needed around all A-stars' within their lifetime to account for their observed prevalence (0.02 * 1 Gyr = 2 Myr * 10 events).  If fewer A-star systems can achieve such events, then the number of events or estimated lifetime of the events would need to increase.   Therefore it is plausible that the prevalence of debris disks around Lambda Boo stars may be related to recent dynamical activity driving a higher rate of accretion and dust production around these stars.

\subsection{Previous ISM Interactions}

It may be feasible that the pollution could have occurred within the past million or so years given the surface mixing time. The stars may have already left an ISM cloud which caused the abundances anomalies presently observed.  The stars in our sample at most move 30 km/s, which over a million years would translate to about 30 pc of movement from their current location.  Given these stars are less then 100 pc away, they will likely have resided entirely within the local bubble out to 150 pc \citep{RL03}.  Thus, they would not have been able to preserve an abundance anomaly if the migrated from a higher density ISM region outside of the local bubble.  There are local ISM clouds within the bubble that may still have intersected the path of the stars, but deconstructing the local ISM and stellar kinematics would require a much more detailed study to rule out this unlikely, but feasible scenario. It maybe that this mechanism plays a greater role in denser ISM regions but for our local sample is not very likely. If the origin was a past ISM interaction, then there wouldn't necessarily be a correlation with an IR-excess like what is observed (see Section \ref{corr_excess}), since stars will have left the ISM cloud that would produce the IR-excess.

\section{Conclusions}

Through the detailed analysis of the PACS images and careful analysis of the two competing models of external accretion, we have shown that the abundance pattern for Lambda Boo stars are likely not caused by ISM accretion.  \textit{Herschel} has offered an improvement in sensitivity and resolution of far-IR excess emission which allows us to confirm the association with the star and resolves the outer radius of the emission.  Together, this information, allows the SED degeneracy to be broken to determine if the excess emission is ISM or debris disk in nature.  

We conclude that the IR excesses seen around our sample of Lambda Boo stars originate from debris disks because:
\begin{itemize}
\item{3/9 of the targeted stars host resolved emission consistent with debris disks.  The resolved emission was mostly symmetric and without a preference towards the direction of proper motion. (see Section \ref{extentemission})}\\
\item{Bow wave models of dust grains 0.1~$\mu$m in size and astrosilcate composition, typical of the ISM, would place the observed dust too close to the star, where its peak emission would be much hotter than what is observed for the sample.  Some stars have ``warm'' components within this temperature range, but this is likely coincidental. (see Section \ref{sedsec} and \ref{equtemp})}\\
\item{Dust around the targeted stars was confined to a radial extent and temperature consistent with dust grains on the order of $\sim$10~$\mu$m in size which are typical of debris disks, regardless of composition.  Furthermore, the emission was inconsistent with 0.1~$\mu$m grains typical of the ISM. (see Section \ref{extentemission} and \ref{equtemp})} \\
\item{Diffuse or extended emission around the stars outside of 150 AU was not found down to a few mJy noise limit.  Background sources were locally confined and separate from the star (see Section \ref{extentemission})}\\
\end{itemize}

The photospheric excess is more likely to arise from a debris disk than an ISM interaction, but the cause of the abundance anomaly in the stars has yet to be identified.  If we favour the hypothesis of a debris disk as the causal relation to the stellar abundance anomaly, then we need to ascertain how the accretion mechanism might function. This requires detailed modeling of the stellar surface to investigate if the required accretion rate is plausible for the debris disks we observe.  It may also be true that the correlation with debris disks is not directly causal but coincidentally related.  Something to consider is that large impacts of planetary bodies or heightened influx of comets could provide the volatile gases for accretion onto the star at a higher rate than PR-drag. In that case, the debris disk may be a symptom of dynamical stirring coinciding with planetary migration, rather than the singular mechanism of accretion. This scenario would also provide an explanation as to why not all debris disk hosts exhibit Lambda Boo characteristics.  For stars of a later spectral type, planetary stirring could still occur but the effect of accretion would be mitigated by a convective envelope, which leads to a cutoff for the Lambda Boo phenomenon in the F-type stars.
 
Future observations which can verify conclusions in this paper are feasible in the near future.  For example, resolving the warm inner components for some of the two component stars in our sample through high contrast imaging would verify the emission is not from disk+ISM interaction systems.  As well as looking for dynamically induced structure. Furthermore, searching for a correlation to exozodi dust around Lambda Boo stars with near-IR interferometers could determine if they are indeed surrounded by hot dust sublimating volatile gas near the surface. The true incidence of IR-excess with Lambda Boo stars will have to be answered by a statistical, spectroscopic study of stars to more rigorously identify Lambda Boo stars with and without debris disk detections to see if there are indeed effects of secondary accretion attributable to debris disks.

\section*{Acknowledgments}
ZHD and BCM acknowledge a Discovery Grant and Accelerator Supplement from the Natural Science and Engineering Research Council of Canada. This work was supported by the European Union through ERC grant number 279973.  We wish to thank the anonymous referee who's comments strengthened this paper.

\footnotesize{
	\bibliographystyle{mn2e}
	\bibliography{ms}
}

\label{lastpage}

\end{document}